\documentstyle[psfig,floats,aps,twocolumn]{revtex}

\def\begeq{\begin{equation}}
\def\endeq{\end{equation}}
\def\beglett{\begin{mathletters}}
\def\endlett{\end{mathletters}}
\def\dtil{\tilde D}
\def\ctil{\tilde c}
\def\ctilinf{\tilde c_\infty}
\def\mtil{\tilde m}
\def\Mtil{\tilde M}
\def\mutil{\tilde\mu}
\def\grad{{\mathbf\nabla}}
\def\lmin{\lambda_{\rm min}}
\def\lmax{\lambda_{\rm max}}
\def\dtmin{\Delta T_{\rm min}}
\def\fsol{f_{\rm sol}}
\def\fliq{f_{\rm liq}}

\begin{document}

\draft\twocolumn[\hsize\textwidth\columnwidth\hsize\csname
@twocolumnfalse\endcsname

\title{Eutectic colony formation: A phase field study}

\author{Mathis Plapp$^{1,2}$ and Alain Karma$^1$}

\address{$^1$Physics Department and Center for Interdisciplinary
             Research on Complex Systems, \\
             Northeastern University, Boston MA 02115 \\
         $^2$Laboratoire de Physique de la Mati{\`e}re Condens{\'e}e,\\
             CNRS UMR No. 7643, Ecole Polytechnique, 91128 Palaiseau, France}

\date{\today}

\maketitle

\begin{abstract}
Eutectic two-phase cells, also known as eutectic colonies,
are commonly observed during the solidification of ternary
alloys when the composition is close to a binary eutectic
valley. In analogy with the solidification cells formed in 
dilute binary alloys, colony formation is triggered by 
a morphological instability of a macroscopically planar
eutectic solidification front due 
to the rejection by both solid phases of
a ternary impurity that diffuses in the liquid.  
Here we develop a phase-field model of 
a binary eutectic with a dilute ternary impurity and we
investigate by dynamical simulations both the initial
linear regime of this instability, and the subsequent highly nonlinear
evolution of the interface that leads to fully developed
two-phase cells with a spacing much larger than the lamellar spacing. 
We find a good overall agreement with our recent linear stability analysis
{\rm [}M.~Plapp and A.~Karma, Phys. Rev. E {\bf 60}, 6865 (1999){\rm ]},
which predicts a destabilization of the front by
long-wavelength modes that may be stationary or oscillatory.
A fine comparison, however, reveals   
that the assumption commonly attributed to Cahn that lamella grow perpendicular to
the envelope of the solidification front is weakly violated in the phase-field 
simulations. We show that, even though weak, this violation
has an important quantitative effect on the stability properties
of the eutectic front.
We also investigate the dynamics of fully developed colonies and find
that the large-scale envelope of the composite eutectic front does not
converge to a steady state, but exhibits cell elimination
and tip-splitting events up to the largest times simulated.  
\end{abstract}

]

\section{Introduction}

Eutectic alloys can form a wealth of different
two-phase patterns during solidification. These
alloys are of interest to metallurgists \cite{KurFis} because
of their low melting points and of the superior mechanical 
properties associated with a fine-scale 
composite microstructure. Moreover, eutectic growth
is a non-trivial example of pattern formation outside
of equilibrium that has attracted the attention of physicists
over the last two decades. 

When the two solid phases ($\alpha$
and $\beta$) of a binary eutectic alloy
have rough interfaces with the liquid, solidification
at or near the eutectic composition typically produces
a spatially periodic array structure consisting of  
lamellar plates of the two phases, or of rods of
the phase with the smaller volume fraction
embedded inside the matrix of the other phase.
Since the pioneering mathematical 
analyses by Hillert \cite{Hillert57} 
and Jackson and Hunt \cite{Jackson66}, which built on
earlier work by Brandt \cite{Brandt45} and Zener \cite{Zener46},
it is well established that
these lamellar and rod morphologies can grow cooperatively 
in steady state for a continuous range of eutectic spacings, 
with both phases helping each other to grow via
the diffusive transport of the two 
chemical components in the liquid
(coupled growth).

In directional solidification experiments, a sample
containing the alloy is pulled at a constant velocity $v_p$
in an externally imposed temperature gradient of
magnitude $G$. In such experiments,
coupled growth typically produces a composite 
front that is perpendicular to the temperature 
gradient, and planar on a scale much larger than 
the lamellar spacing $\lambda$ (defined as the 
width of the basic spatially repeating unit consisting
of one $\alpha$-lamella and one $\beta$-lamella).
Analytical \cite{Datye81} and numerical 
\cite{Karma96} studies of
the morphological stability of lamellar eutectics,
as well as detailed experiments in a transparent organic
system \cite{Ginibre97}, have shown that the stable range of 
lamellar spacings is restricted by a long-wavelength 
instability leading to local lamellar termination
at small $\lambda$, and short-wavelength 
oscillatory instabilities at large $\lambda$.
These studies clearly demonstrate that a large-scale
morphological instability of the composite front
does not occur in a binary eutectic alloy.

\begin{figure}
\centerline{
  \psfig{file=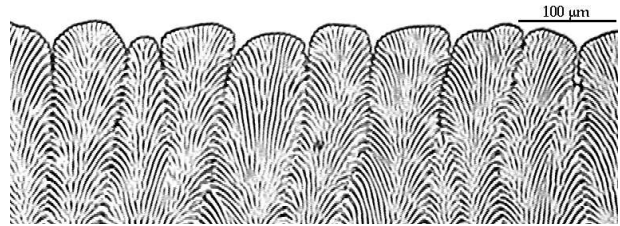,width=.45\textwidth}}
\medskip
\caption{Eutectic colonies in a thin sample of the transparent
organic eutectic alloy $\rm CBr_4-C_2Cl_6$, doped with a
small amount of the ternary impurity naphtalene (from
Ref. \protect\cite{Akamatsu00}).}
\label{figexp}
\end{figure}
This picture is consistent with 
the experimental observation that such a 
morphological instability occurs only 
when a small quantity of a ternary impurity   
is present, and when $v_p$ exceeds a critical value
\cite{Weart58,Chilton61,Kraft61,Hunt66,Gruzleski68,Rumball68,Rohatgi69,Bullock71,Rinaldi72,Akamatsu00,Han98}. 
In a nonlinear regime, this instability gives rise
to the formation of two-phase solidification cells, also called 
eutectic colonies, with a typical width much larger than
$\lambda$. A typical example of such cells is shown in 
Fig. \ref{figexp}.

Experimental measurements to date 
\cite{Weart58,Chilton61,Kraft61,Hunt66,Gruzleski68,Rumball68,Rohatgi69,Bullock71,Rinaldi72,Akamatsu00,Han98}
have consistently shown that
the onset of colony formation can be relatively well
described by a simple constitutional supercooling criterion
with respect to the ternary impurity \cite{Tiller58,McCartney80}, 
which predicts that instability occurs when $G/v_p$
falls below a critical value. This 
suggests that this instability may be 
qualitatively similar to 
the well-known Mullins-Sekerka instability  
of a monophase front
during directional solidification of a dilute binary 
alloy \cite{Mullins64}.
In a recent linear stability analysis of
a sharp interface model \cite{Plapp99}, however,
we showed that the morphological
instability of a composite front in the presence
of a dilute ternary impurity differs fundamentally
from the instability of a monophase front, even though
the onset of both instabilities is well predicted by  
constitutional supercooling. This analysis
was based on the same procedure
used previously by Datye and Langer    
\cite{Datye81} to analyze the stability of
binary lamellar eutectics, where the basic
degrees of freedom are the coordinates of the $\alpha$-$\beta$-liquid
trijunctions. Our main finding was that the  
amplification of linear perturbations of 
the composite front can be either
steady or oscillatory for experimentally relevant
control parameters,
in contrast to the classical Mullins-Sekerka instability
where finite-wavelength perturbations  
are amplified in a non-oscillatory way.

Furthermore, in Ref. \cite{Plapp99}, 
we developed an ``effective monophase front'' 
formulation of the dynamics of the composite
interface that shed light on the origin of this difference.
We showed that the long-wavelength dynamics of 
the envelope of the composite front is governed  
by a free-boundary problem with boundary conditions for the
concentration of the diffusing ternary
impurity on the effective front that can be obtained by averaging 
over the properties of the two solid phases. As a self-consistency check,
we also showed that, when the wavelength of the perturbation is 
much larger than $\lambda$, the linear stability analysis 
of this free-boundary problem gives
identical results to the full stability calculation 
expressed in terms of the trijunction coordinates.

Not surprisingly,
this free-boundary problem turns out to be very similar
to the one governing a ``true'' monophase front in
a dilute binary alloy. The non-trivial difference, however,
is that the local lamellar spacing, which appears in the
boundary condition for the ternary impurity
on the front, constitutes an additional ``internal
degree of freedom'' of the front that modifies its
stability properties, and gives rise to the oscillatory modes.
Physically, this reflects the fact that the
local temperature of the front depends on the local lamellar spacing
$\lambda$ and that, in turn, the time rate of change of $\lambda$ depends
on the shape of the front because of the geometrical constraints
imposed by the equilibrium conditions for the angles between
interfaces at the trijunctions (Young's conditions).

In a recent experimental study of a transparent organic
model alloy, oscillatory patterns compatible with the results
of our linear stability analysis were indeed observed \cite{Akamatsu00}.
The same study, however, also revealed a wealth of other possible
structures that can be associated with the instability of a
planar front, and in particular localized two-phase fingers
that may appear in an early stage of the morphological instability.

The two main goals of the present study are to check the
validity of our previous linear stability analysis \cite{Plapp99}
by direct simulation of the fundamental equations of motion, 
and to investigate the nonlinear regime of colony formation.
For this purpose, we develop a phase-field model for
the directional solidification
of a eutectic alloy with a dilute ternary impurity.
Simulations of this model
enable us to characterize quantitatively the amplification
and decay of linear
perturbations of the composite front
and to study the complex interface dynamics leading to
the formation of well-developed colonies. 

The phase-field method is by now a well-established
technique for simulating solidification 
patterns \cite{Langer86,Fix83,Collins85,Wheeler92,Karma98,Karma01}. 
In particular, it has already been applied to the investigation of 
multiphase solidification in eutectic and peritectic alloys 
\cite{Karma94,Elder94,Wheeler96,Tiaden98,Nestler00,Lo01}. 
The advantage of this method with respect to the boundary 
integral formalism used previously to perform detailed 
simulations of eutectic growth structures \cite{Karma96} is 
that ternary impurities can easily be included. Furthermore,
the phase field method is able to handle
automatically dramatic changes in the interface morphology
such as lamella termination and creation, which are
difficult to implement in the boundary integral approach.

The phase-field model presented in this paper is specifically
designed for computational efficiency and therefore makes some
simplifying assumptions. In particular, we use a generic
eutectic phase diagram that is symmetric with respect to
the exchange of the two solid phases, and we neglect 
crystallographic effects such as the anisotropy 
of the solid-liquid and solid-solid
interfacial energies. The computational
effort required to simulate fully developed
colonies is nonetheless considerable
since the two-phase cell spacing
is one order of magnitude larger than 
$\lambda$. For this reason, the
largest simulations of such
structures were carried out on a multi-processor CRAY T3E  
and took the equivalent of a few thousand hours of
single-processor workstation time.
 
The simulation results are found to be in good overall
agreement with our sharp-interface linear stability
analysis for compositions close to the 
eutectic point, where the two solid phases have approximately equal 
volume fractions. We observe, indeed, the predicted 
large-scale oscillatory structures.
Quantitatively, however, the simulated growth rates differ
from the predicted ones. A careful analysis of our simulation
results, extrapolated to the limit of vanishing 
thickness of the diffuse interfaces,
allows us to pinpoint the origin of this discrepancy.
In particular, our stability analysis uses the assumption that the 
lamellae grow normal to the large-scale growth front. 
This assumption is commonly attributed to Cahn and was also used
previously by Datye and Langer \cite{Datye81} for their
linear stability analysis of lamellar eutectics in binary alloys. 
We find that, in our simulations, this rule is slightly violated. 
Hence, the stability analysis correctly describes all the qualitative
features of the instability, but would have to be extended to 
include this effect in order to become quantitatively accurate.
This violation also has important consequences for the stability
of binary eutectics that will be discussed elsewhere.

The linear instability of the planar front is followed by
a nonlinear transient that leads to the formation of
fully developed colonies. The nature of the transient
depends on the composition. In simulations carried out
at the eutectic point, the long-wavelength modes grow
until the front becomes wavy and the first lamella
terminations occur in the concave parts.
Subsequently, the grooves deepen and the tips grow
ahead of the front, such that the initial wavelength
of the colonies corresponds to the linear mode that
dominates the stability spectrum. In contrast, for
off-eutectic compositions, the linear regime is much
shorter, and localized two-phase fingers centered
around a thin lamella of the minority phase grow rapidly 
ahead of the front and develop into colonies later on.

Finally, once formed, the colonies have a quite well-defined
average size and shape at both eutectic
and off-eutectic compositions. However, the front does not
settle down into a true steady state, but exhibits 
tip-splitting and cell elimination events, not
unlike the monophase front of a dilute alloy in
the absence of interfacial anisotropy \cite{Akaetal,Kopetal}.

The remainder of this paper is organized as follows. In the next
section, we introduce the phase-field model 
and analyze its sharp-interface limit. 
In section III, we present simulation
results for stable steady-state
lamellar growth that are used to test our model. 
Section IV contains a brief review of our sharp-interface
linear stability analysis \cite{Plapp99}, 
and a detailed comparison between the analytical 
and numerical results concerning the linear 
stability of the eutectic front.
Section V is devoted to the simulations of well-developed
colony structures in a nonlinear regime. 
Finally, conclusions and an outlook 
for future work are given in section VI.

\section{Phase-field equations and sharp-interface limit}

We consider directional solidification of thin samples, 
as used in many experimental studies of pattern formation
during solidification 
\cite{Ginibre97,Hunt66,Akamatsu00}.
This allows us to treat the problem as essentially
two-dimensional and to neglect convection in the liquid.
Furthermore, we assume that the rejection of latent heat during
solidification does not appreciably modify the temperature
field created by the experimental setup (frozen temperature
approximation), and hence that growth is limited by diffusion
of the chemical constituents.

We are interested in the behaviour of a ternary alloy
close to a binary eutectic trough in the phase diagram.
Specifically, we will consider a very low
concentration of the third component,
which can then be regarded as a dilute impurity.
This allows us to neglect various cross-coupling
terms between the ternary impurity and the
components of the binary eutectic.
In addition, we are more interested in
generic aspects of two-phase cell formation than in
modeling a specific material. Hence, we study a
model eutectic alloy that has a symmetric phase diagram. 
This simplifies the setup of the phase-field model. 

The principles of the phase-field method have been 
described in detail in numerous publications 
\cite{Langer86,Fix83,Collins85,Wheeler92,Karma98,Karma01,Karma94,Elder94,Wheeler96,Tiaden98,Nestler00,Lo01}.
The idea is to distinguish between the different thermodynamic
phases with the help of one or several scalar fields, the
phase fields, that have fixed values in the bulk phases
and vary continuously across smooth and diffuse interfaces.
A free energy functional suitable for the problem at hand
is then constructed, and the equations of motion for the
fields are written in variational form. By now, 
various phase-field models for alloy solidification are available
\cite{Wheeler92,Karma01,Karma94,Elder94,Wheeler96,Tiaden98,Nestler00,Lo01}. 
In particular, much effort was spent to develop
a thermodynamically consistent approach and to base the
free energy functional on ideal or regular solution 
models \cite{Wheeler92,Wheeler96,Nestler00}.
In contrast, we are interested here mainly in the phase-field
model as a computational tool. We will therefore use a strongly
simplified model that is chosen for its computational efficiency,
with the minimum of ingredients necessary to reproduce the
main features of eutectic solidification with a ternary impurity.
The parameters of the model are related to physical quantities
by performing a sharp-interface limit.

We choose as the set of dynamical field variables
the concentration (in molar fraction) $c(x,z,t)$ 
of one of the components of the binary eutectic, the concentration 
$\ctil(x,z,t)$ of impurities, and a single phase field $\phi(x,z,t)$
that distinguishes between solid and liquid.
To simplify the construction of the free energy
functional, we define the scaled concentration $u$ by
\begeq
u(x,z,t) = {c(x,z,t)-c_E\over\left(c_\beta-c_\alpha\right)/2},
\label{uscaled}
\endeq
where $c_E$, $c_\alpha$, and $c_\beta$ are the compositions of
the liquid and the two solid phases in the pure binary eutectic
at the eutectic temperature $T_E$ \cite{rem0}. For a symmetric 
phase diagram, the scaled compositions of the two solids at $T_E$
are $u=\pm 1$.

Building on a previous phase-field model for a binary 
eutectic \cite{Karma94}, we take the (dimensionless) 
free energy functional \cite{rem1} of the form
\begeq
F = \int_V dV\left[ {W_u^2\over 2} {(\grad u)}^2 + 
           {W_\phi^2\over 2} {(\grad \phi)}^2 +
            f(\phi,u,\ctil,T)\right],
\endeq
where $V$ is the volume of the two-phase system.
The dimensionless free energy density $f(\phi,u,\ctil,T)$
must have three local minima to account for the three
possible phases (liquid, $\alpha$ solid, and $\beta$ solid),
separated by potential barriers. We use the phase field
to distinguish between solid and liquid, and the scaled
concentration field to distinguish between the two solids.
The gradient terms force the fields to vary continuously
between the bulk equilibrium values and hence create interfaces
of a characteristic thickness of order $W_u$ (solid-solid interface) 
and $W_\phi$ (solid-liquid interfaces). In general, there
should also be a gradient term for the ternary impurity.
However, we may omit this term for simplicity
since $\ctil$ has no indicator function, but is slaved
to the other fields; that is, for specified phase field $\phi$,
concentration $u$ and temperature $T$, the equilibrium value
of $\ctil$ is known.

A convenient choice for the free energy density is
\begin{eqnarray} 
f(\phi,u,\ctil,T)& = & -{\phi^2\over 4}+{\phi^4\over 8} + {1+h(\phi)\over 2}\fsol(u,\ctil,T)
  \nonumber \\
                 &   & \quad\mbox{}    +{1-h(\phi)\over 2}\fliq(u,\ctil,T).
\label{fdecomp}
\end{eqnarray}
Here, $\fsol$ and $\fliq$ are the bulk free energy densities
in the solid and the liquid, respectively, and
\begin{equation} 
h(\phi)={3\over 2}\left(\phi-{\phi^3\over 3}\right)
\end{equation}
is an interpolation function.
The first two terms in Eq. (\ref{fdecomp}) generate a 
double well potential for $\phi$ with minima at $\phi=\pm 1$. 
Since $h(\pm 1)=\pm 1$, $f(1,u,\ctil,T)=\fsol(u,\ctil,T)$ and
$f(-1,u,\ctil,T)=\fliq(u,\ctil,T)$, such that $\phi=+1$
corresponds to the solid and $\phi=-1$ to the liquid.
Moreover, since $h'(\pm 1)=0$, the equilibrium values
of $\phi$, given by the solutions of $df/d\phi = 0$,
always remain at $\phi=\pm 1$, independently of the
values of $\fsol$ and $\fliq$.

For $\fsol$ and $\fliq$, we take
\begin{eqnarray}
\fliq(u,\ctil,T) & = & u^2/2 + b\ctil\ln\ctil - \epsilon_l\ctil, \\
\fsol(u,\ctil,T) & = & a\left(u^2-1\right)^2 + b \ctil\ln\ctil -
    \epsilon_s\ctil - \alpha\Delta T/T_E,
\label{fsoleq}
\end{eqnarray}
where $\Delta T = T_E-T$ is the undercooling with respect 
to the binary eutectic point, and $a$, $b$, $\epsilon_s$, $\epsilon_l$,
and $\alpha$ are constants that will be related to physical
parameters by the construction of the phase diagram. This
choice is motivated by the following considerations. Since
there are two solid phases, $\fsol$ must have a double-well
structure in $u$; in contrast, $\fliq$ has a single well.
At the eutectic temperature and without impurities
($\ctil_to 0$), all three phases must have
the same free energy; for $T>T_E$ ($T<T_E$), the liquid minimum
must be below (above) the solid minima. This is conveniently
achieved by the last term on the right hand side of
Eq. (\ref{fsoleq}) that simply shifts $\fsol$ with respect
to $\fliq$; formally, $\alpha$ is equivalent to the
latent heat. 

The impurity terms have a
form that is equivalent to the dilute limit of a regular
solution model. Indeed, the terms containing $\ctil$ correspond to
the dilute approximations for the entropy of mixing and the
energy cost of the impurities, respectively, with $\epsilon_\nu$
representing the difference in bond energies upon replacing
a ``solvent'' atom by an impurity in phase $\nu$. The constant
$b$, which sets the energy scale, should formally be proportional
to the temperature. Since we are, however, only interested in a
narrow temperature range around $T_E$, we simply use a constant.

The various coefficients can be related to physical quantities
through the construction of the phase diagram. The conditions
for thermodynamical equilibrium between two distinct phases
are (i) equal chemical potentials for the eutectic components,
(ii) equal chemical potentials for the ternary impurity, and
(iii) equal grand potential, i.e.
\begin{eqnarray}
\mu_s \equiv \left.\partial\fsol/\partial u\right|_{u_s} & = &
    \mu_l \equiv \left.\partial\fliq/\partial u\right|_{u_l} ,
 \label{euteq}\\
\mutil_s \equiv \left.\partial\fsol/\partial\ctil\right|_{\ctil_s} & = &
    \mutil_l \equiv \left.\partial\fliq/\partial\ctil\right|_{\ctil_l} ,
 \label{impeq} \\
\Omega_s \equiv \fsol-\mu_s u_s - \mutil_s \ctil_s & = &
    \Omega_l \equiv \fliq-\mu_l u_l - \mutil_l \ctil_l,
 \label{omeq}
\end{eqnarray}
where $u_\nu$ and $\ctil_\nu$, $\nu = s,l$, denote the equilibrium
concentrations in solid and liquid. These conditions can be
geometrically described as a ``common tangent plane'' to the
free energy surface, analogous to the well-known double-tangent
construction for binary alloys. From Eq. (\ref{impeq}), we get
at once the standard partition relation for a dilute alloy,
\begeq
\ctil_s = K \ctil_l,
\label{partition}
\endeq
with a partition coefficient given by
\begeq
K = \exp[-(\epsilon_s-\epsilon_l)/b].
\label{partitioncoeff}
\endeq
Next, from Eq. (\ref{euteq}), under the assumption $\Delta T/T_E \ll 1$
(i.e. for temperatures close to the eutectic temperature),
we have $u_l \ll 1$ and $u_s \approx \pm 1$, and we get
\begeq
u_s = {u_l\over 8a} \pm 1,
\label{gaps}
\endeq
where the two signs correspond to the two distinct 
solid-liquid equilibria. Finally, using Eqs.
(\ref{omeq}), (\ref{partition}), and (\ref{gaps}),
we obtain
\begeq
\alpha\Delta T/T_E = b(1-K)\ctil_l \pm u_l.
\endeq
Using the definition of $\Delta T$, this can be rewritten as
\begeq
T=T_E \pm m_u u_l - \mtil\ctil_l,
\endeq
where $m_u$ and $\mtil$ are the magnitudes of the liquidus 
slopes for the eutectic components and the impurity, respectively,
\begin{eqnarray}
m_u & = & \left|dT\over du_l\right| = {T_E\over\alpha},
\label{liquiduseut} \\
\mtil & = & \left|dT\over d\ctil_l\right| = 
      {b(1-K)T_E\over\alpha}.
\label{liquidusimp}
\end{eqnarray}
Note that the scaled liquidus slope $m_u$ can be related
to the ``true'' liquidus slope $m$ in the phase diagram
with the help of Eq. (\ref{uscaled}),
\begeq
m_u= m (c_\beta-c_\alpha)/2.
\endeq
The parameter $a$ controls the ratio of the liquidus
and solidus slopes in the eutectic phase diagram;
for simplicity, we will fix in the following $a=1/8$, 
which gives a concentration jump across the interface
that is independent of temperature (parallel liquidus
and solidus lines). The parameter $b$, together
with the partition coefficient $K$, fixes the
ratio of eutectic and impurity liquidus slopes,
$\mtil/m_u=b(1-K)$. 

The equations of motion for the three fields are
\begin{eqnarray}
\tau \partial_t \phi(x,z,t) & = & - {\delta F\over \delta \phi(x,z,t)}, 
  \label{phidyn} \\
\partial_t u(x,z,t) & = & \grad \left( M(\phi,u,\ctil) \grad
       {\delta F\over \delta u(x,z,t)}\right), \\
\partial_t \ctil(x,z,t) & = & \grad \left( \Mtil(\phi,u,\ctil) \grad
       {\delta F\over \delta \ctil(x,z,t)}\right),
\end{eqnarray}
where $\delta F/\delta\cdot$ denotes the functional derivative
with respect to the field $\cdot$, $\tau$ is a (microscopic)
relaxation time, and $M$ and $\Mtil$ are the
mobilities of the eutectic component and the ternary impurity,
respectively. These variational forms reflect the fact 
that the two concentrations are conserved fields, whereas 
the phase field can be seen as a non-conserved order parameter. 
The non-conserved phase-field simply relaxes
towards its local equilibrium value. Indeed, by inserting 
Eq. (\ref{fdecomp}) into Eq. (\ref{phidyn}) we obtain
\begeq
\tau\partial_t\phi = W_\phi \nabla^2 \phi + \phi/2-\phi^3/2 + 
  h'(\phi)(\fliq-\fsol).
\endeq
The last term on the right hand side always drives the phase 
field to the value that corresponds to the lower local free 
energy density (we recall that $h'>0$ and that $\phi=1$ 
corresponds to the solid).

The definition of the model is completed by the specification
of the mobility functions $M(\phi,u,\ctil)$ and 
$\Mtil(\phi,u,\ctil)$. The dependence of $M$ and $\Mtil$ on the 
phase field and the compositions allows us to obtain the 
desired diffusivities in the bulk phases.
We want to simulate a one-sided
model (i.e. vanishing diffusivity in the solid) with
constant diffusivities for eutectic components
and impurities in the liquid. This can be achieved by choosing
\beglett
\label{mobilities}
\begin{eqnarray}
M(\phi,u,\ctil) & = & D\left[1-
\left(\frac{1+\phi}{2}\right)^n\right],  \\
\Mtil(\phi,u,\ctil) & = & \dtil \left[1-
\left(\frac{1+\phi}{2}\right)^n\right] \ctil,
\end{eqnarray}
\endlett
where $D$ and $\dtil$ are the diffusion constants.
Indeed, from the equations of motion, we get that
in the liquid ($\phi\equiv-1$)
\begin{eqnarray}
\partial_t u & = & \nabla\left[ M \left(
      {\partial^2\fliq\over\partial u^2}\nabla u + 
        W_u^2\nabla(\nabla^2 u)\right)\right], \\
\partial_t \ctil & = & \nabla \left( \Mtil
      {\partial^2\fliq\over\partial \ctil^2} \nabla\ctil\right).
\end{eqnarray}
In the first equation, we can neglect the term $W_u^2\nabla(\nabla^2u)$
in the brackets on the right-hand side,
since the diffusion pattern forms on a scale much larger than
$W_u$, and hence this term is small compared to
$(\partial^2\fliq/\partial c^2) \nabla u$. Using the
expressions for the mobilities and $\fliq$, we obtain the
desired result in the liquid,
\begin{eqnarray}
\partial_t u & = & D \nabla^2 u, \\
\partial_t \ctil & = & \dtil \nabla^2 \ctil.
\end{eqnarray}
The exponent $n$ in the mobility plays a role only in the
interfacial region where the phase field varies, and changing
its value modifies the interface kinetics. This will be
discussed in more detail below.

When the thickness of the diffuse interfaces is much smaller
than all other physical length scales, and in particular
the lamellar spacing $\lambda$, the above phase-field equations
can be related to the more conventional sharp-interface
equations of the macroscopic models of solidification by
the technique of matched asymptotic expansions. This
procedure has been detailed in several publications for
models that are similar to ours, and hence we will only
outline the results. 
Each solid has to reject its minority component and the ternary
impurity into the liquid in order to grow. Since the concentrations
are locally conserved quantities, mass balance at the interface 
implies that they obey boundary conditions of Stefan type at
the moving boundary, i.e.
\begin{eqnarray}
      -D \partial_n u & = & v_n (u_l-u_s),\\
      -\dtil \partial_n \ctil & = & v_n (1-K)\ctil_l.
\label{stefan}
\end{eqnarray}
Here, $v_n$ and $\partial_n$ are the normal velocity of the
interface and the derivative normal to the interface, and
$u_l$, $u_s$, and $\ctil_l$ are the values of the concentrations
at the liquid and solid sides of the interface, respectively.
These equations are valid for both solid-liquid interfaces; note
that on the $\alpha$-liquid interface, $u_l-u_s>0$, whereas
on the $\beta$-liquid interface, $u_l-u_s<0$. The concentrations
at the interface are related to temperature, shape, and speed
of the interface by a generalized Gibbs-Thomson condition,
\begeq
T = T_E \mp m_u u_l - \mtil \ctil_l - \Gamma {\cal K} - v_n/\mu_k,
\label{githo}
\endeq
where the upper (lower) sign is for the $\alpha$ ($\beta$) 
phase, the liquidus slopes $m_u$ and $\mtil$ are given by Eqs. 
(\ref{liquiduseut}) and (\ref{liquidusimp}), $\Gamma$ is
the Gibbs-Thomson constant, $\cal K$ is the local curvature
of the interface, $\mu_k$ is the linear kinetic 
coefficient, and the concentrations on the solid side are
linked to those on the liquid side via Eqs. (\ref{partition}) 
and (\ref{gaps}). Without the last (kinetic) term, Eq. (\ref{githo})
is a statement of local equilibrium at the interface, including
capillary effects. The Gibbs-Thomson constant $\Gamma$ is given in
physical units by $\Gamma=\gamma_{sl}T_E/L$, where $\gamma_{sl}$
is the surface tension of the solid-liquid interface, and $L$
is the latent heat of melting. In the units of our model, this becomes
\begeq
\Gamma = \gamma_{sl} T_E/ \alpha.
\endeq
The surface tension $\gamma_{sl}$ is obtained as in Ref. \cite{Karma94} 
by first solving numerically the one-dimensional stationary versions
of the equations of motion to obtain the interface profiles,
and then computing the excess free energy per unit surface
by inserting the profiles into the free energy functional.
Note that, since the free energy density is taken dimensionless here,
surface tensions have units of length. The solid-solid
surface tension can be calculated analytically since along
the $\alpha\beta$-interface $\phi\equiv 1$, and we obtain
$\gamma_{ss}=(2/3)W_u$. 

From the surface tensions, we can
determine the contact angle $\theta$, that is, the angle between 
the horizontal and the solid-liquid interfaces at a trijunction 
point where the solid-solid interface is vertical. Using 
Young's condition of mechanical equilibrium, we get
\begeq
\sin \theta = {\gamma_{ss}\over 2 \gamma_{sl}}.
\label{young}
\endeq

We have not explicitly calculated the value of the kinetic
coefficient $\mu_k$ that appears in the last term of Eq. (\ref{githo}). 
This would require to compute several integrals in the 
coupled variables $u$ and $\phi$ through the solid-liquid 
interface (see Ref. \cite{Karma98} for more details in a 
simpler case), which can only be done numerically. 
Furthermore, we have neglected in the above analysis
other non-equilibrium effects, and in particular solute
trapping \cite{Aziz82} that is generally present in 
phase-field models for alloy solidification \cite{Karma01,Wheeler93}. 
It is known that solute trapping modifies
the compositions on both the liquid and the solid sides
of a moving interface. This generates correction terms
both in the Gibbs-Thomson condition and the mass balance
relations, Eqs. (\ref{stefan}) and (\ref{githo}). However,
these corrections are proportional to the interface
velocity, and are expected to be small for the range
of solidification speeds used in our present simulations.
Indeed, as we will see below, the non-equilibrium
effects are not entirely negligible; however, 
they are not important enough to justify a detailed
analysis that would be quite involved \cite{Karma01}.

\section{Lamellar steady states}

We chose as a testing ground for our model the simulation
of lamellar steady-state solutions. This has the additional
benefit of providing us with the initial configurations needed 
for the simulations of large-scale arrays described below.
In the laboratory frame, the sample is pulled with velocity 
$v_p$ in a constant temperature gradient $G$ along the $z$-axis.
This means that in the sample frame, the isotherms move
towards the positive $z$ direction with velocity $v_p$. Consequently,
the temperature at a given point $(x,z)$ of the sample is
\begeq
T(z,t) = T_E + Gz - v_pt,
\label{temperature}
\endeq
where we have chosen the origin of the $z$-axis at the
eutectic isotherm for $t=0$.

The equations of motion were simulated by an explicit
Euler algorithm with timestep $\Delta t$ on a simple
square grid of spacing $\Delta x$ using standard 
finite-difference formulae.
For simplicity, we chose $W_u = W_\phi \equiv W$.
In the following, unless otherwise stated, lengths will
be measured in units of $W$, time in units of $\tau$, and
temperatures in units of $T_E$. We chose $D=\dtil=1$,
$\alpha=1$, $a=1/8$ (parallel eutectic solidus and 
liquidus lines), $G=0.001$, $v_p$ between $0.005$ 
and $0.02$, and various values of $b$ and $K$,
with $\epsilon_l=0$ and $\epsilon_s=-b\ln K$. 
Since the equation for the composition $u$ is of 
fourth order, the critical timestep for the occurrence
of numerical instabilities scales as $\Delta x^4$.
The allowed grid spacing $\Delta x$, however, is limited
by the requirement that the smooth interfaces be
sufficiently well resolved to avoid strong numerical
anisotropies and lattice pinning. We found that $\Delta x = 1$
and $\Delta t = 0.025$ provided a good compromise between
efficiency and accuracy.

The simulations were started with a single pair of 
flat lamellae in contact with the liquid in a box
of lateral size $\lambda$.
The concentrations were set to the equilibrium values 
in each phase. For the subsequent evolution, periodic 
boundary conditions were used in the direction parallel 
to the temperature gradient, while the concentrations
in the liquid were kept at fixed values $u_\infty$ and $\ctil_\infty$
at the upper end of the simulation box. At the lower (solid) end,
no boundary conditions are needed since the fields
do not evolve. During the runs, the simulation box
was periodically shifted to follow the interface.
Convergence to the steady-state solution was checked by computing 
the average change of the phase field in the moving frame 
during the advance of the isotherms by one lattice spacing.
Furthermore, the interface shapes (given by the level set 
$\phi=0$ for the solid-liquid interface and by $u=0$ in the
solid, that is for $\phi>0$, for the solid-solid interfaces) 
are extracted by interpolation of the fields between the
lattice points. This procedure yields a resolution far superior 
to the grid spacing. The average undercooling of the 
interface is then
\begeq
\Delta T(t) = T_E - T_{int}(t) = 
  - G \left({1\over\lambda}\int_0^\lambda \zeta(x,t) dx \, - v_p t\right), 
\endeq
where $\zeta(x,t)$ is the $z$-position of the extracted 
solid-liquid interface as a function of $x$ at time $t$.
The simulations were stopped when the undercooling
was to within $10^{-4}$ of its extrapolated final value.

We first discuss the special case of a pure
(binary) eutectic at the eutectic composition,
$u_\infty=\ctil_\infty=0$ (note that we omit the impurity terms 
in the free energy and the equation of motion for the impurities 
when $\ctil_\infty=0$). For our symmetric phase 
diagram, there is no global diffusion boundary layer
in this case, and the diffusion field in the liquid
decays exponentially on a scale of $\lambda$.
Hence, a box length parallel to the temperature
gradient of about five times $\lambda$ was sufficient to
obtain results that are independent of the box size. 
The interface relaxes exponentially to its steady state,
with relaxation times of order $\lambda^2/D$;
on a typical modern workstation, the convergence
takes a few hours. 

In contrast, for $u_\infty\neq 0$ and/or
$\ctil_\infty\neq 0$, solute redistribution leads to a boundary
layer of thickness $l_D=D/v_p$, much larger than $\lambda$.
Hence, box sizes along the growth direction of several
times $l_D$ have to be used; in addition, the interface 
position now follows a damped oscillation with
exponential envelope and decay times of the order $D/v_p^2$,
much larger than $\lambda^2/D$. As a result, when a boundary
layer is present the convergence of a run takes several
days of CPU time. The oscillatory relaxation of the interface 
position is compatible with the Mullins-Sekerka dispersion 
relation at zero wave vector that predicts a complex decay rate.

Let us compare our results to the well-known Jackson-Hunt 
(JH) relation between lamellar spacing and interface 
undercooling \cite{Jackson66}, generalized
to include the effect of the ternary impurities,
\begeq
\Delta T(\lambda) = \Delta T_{JH}(\lambda) + \mtil\ctilinf/K,
\label{undercooling}
\endeq
\begeq
\Delta T_{JH} = {1\over 2}\dtmin
    \left({\lambda\over \lmin}+{\lmin\over\lambda}\right).
\label{jhunt}
\endeq
The curve $\Delta T$ versus $\lambda$ exhibits
a minimum at a spacing $\lmin$, where
\begeq
\dtmin= {2m_u\Delta u\over \eta(1-\eta)}
   \sqrt{\Gamma\sin\theta P(\eta)v_p\over2Dm_u\Delta u},
\label{deltatmin}
\endeq
\begeq
\lmin= \sqrt{2\Gamma\sin\theta D\over m_u\Delta u v_p P(\eta)}.
\label{lambdamin}
\endeq
Here, $\Delta u=u_\beta-u_\alpha=2$ is the concentration
difference between the two solids at the eutectic temperature,
$\eta=(u_\infty-u_\alpha)/\Delta u$
is the volume fraction of $\beta$-phase in the solid,
$P(\eta)=\sum_{n=1}^\infty \sin^2(\pi\eta n)/(\pi n)^3$,
and $\theta$ is the contact angle defined by Eq. (\ref{young}).
For $W_u=W_\phi=1$, we obtain numerically $\gamma_{sl}=1.04$,
which together with $\gamma_{ss}=2/3$ gives $\theta\approx 19^\circ$.
Note that Eqs. (\ref{deltatmin}) and (\ref{lambdamin}) are valid 
only for our choice of a symmetric phase diagram; see Ref. \cite{Plapp99} 
for a discussion of the general case.
It should be kept in mind that the JH theory is approximate
since it uses a flat interface to calculate the diffusion
field. Nevertheless, it has been shown by boundary
integral simulations \cite{Karma96} that the error
is small for small contact angles $\theta$ and close to 
the spacing $\lmin$, such that it can be used as a 
semi-quantitative test for our phase-field model.

\begin{figure}
\centerline{
  \psfig{file=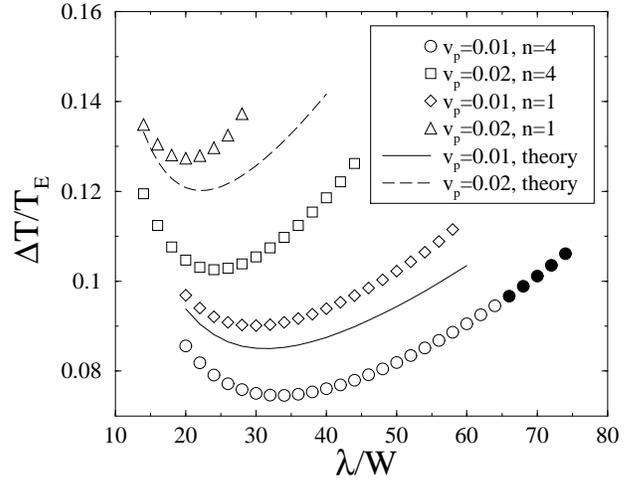,width=.45\textwidth}}
\medskip
\caption{Average interfacial undercooling versus
lamellar spacing for several values of the pulling 
speed $v_p$ and the mobility exponent $n$. Lines: prediction
of the Jackson-Hunt theory, Eq. (\ref{jhunt});
symbols: simulation results. Filled circles correspond
to steady states that are unstable with respect 
to 1-$\lambda$ oscillations.}
\label{figjhunt}
\end{figure}
We computed the interface undercooling in our
model for various pulling speeds and two different
values of the mobility exponent $n$ in Eqs. (\ref{mobilities})
of the mobility functions. Let us first discuss the results
for $n=1$, which corresponds to the simplest form of
the mobility that has been widely used before. The 
simulated undercoolings are slightly higher than
the JH prediction, but the overall shape of the 
curve is perfectly reproduced. The difference
can be attributed to the non-equilibrium effects
(interface kinetics, solute trapping)
present in the phase-field model, but neglected
in the JH theory. Indeed, the differences between
our simulations and the JH prediction are larger 
for higher $v_p$. Furthermore, we obtained
$\lmin$ and $\dtmin$ by fitting our simulation results 
to Eq. (\ref{jhunt}) and found that the scaling relation 
$\lmin^2v_p={\rm const.}$ that can be derived
from Eq. (\ref{lambdamin}) is well satisfied.
Regarding the impurity contribution to 
Eq.~(\ref{undercooling}), we conducted simulations
for various impurity concentrations, impurity 
liquidus slopes and partition coefficients and 
found good agreement with the predicted behavior.
In particular, we verified that the spacing $\lmin$
was not appreciably modified by the addition of
impurities.

The range of lamellar spacings that can be simulated
is limited by two effects that are intrinsic to our model.
For spacings smaller than $\sim 16W$ ($\sim 8W$ for each
individual lamella), the diffuse interfaces at the two sides 
of a lamella start to overlap, which leads to strong corrections
to the sharp-interface limit and ultimately to lamella
elimination. 

For too large spacings, in turn,
new lamellae of the opposite phase nucleate in
the centers of the initial lamellae, leading to a
lamellar array with one third of the initial spacing.
This is the result of a ``spinodal decomposition'' that
takes place in the interface. Indeed, the equation for the
composition in the solid far from the interface is exactly
the classical Cahn-Hilliard equation \cite{Cahn58}, which is known to 
exhibit phase separation without nucleation in a composition
range where the free energy density has a negative curvature 
($\partial^2 \fsol/\partial c^2 < 0$). Far inside 
the solid, this has no importance here because the mobility 
is zero and hence no dynamics takes place. Well within the 
liquid, there is no unstable concentration range since
the liquid free energy has a single-well structure.
But in the diffuse interface, new
domains may form when the concentration falls within
the unstable range. According to the JH theory, the
deviations of the concentration from the equilibrium
value at the interface scale as ${\rm Pe}\,\lambda$, where 
the P\'eclet number ${\rm Pe}=\lambda/l_D=\lambda v_p/D$; 
hence, the maximum spacing $\lmax$ that can be simulated 
before nucleation sets in increases as $v_p$ decreases. 
Indeed, we find $\lmax/W \sim 28$ for $v_p=0.02$ and
$\lmax/W\sim 58$ for $v_p$=0.01.

It seems useful at this point to comment on the implications 
of these limitations for the choice of the computational
parameters for large-scale simulations. The range 
of initial lamellar spacings of
interest for the present study ranges from $\lmin$ to 
about $1.5\,\lmin$. Since we want to simulate the linear
instability of a lamellar front, which may involve
considerable variations of the local lamellar spacing,
the model should work for a sizeable range of spacings,
say at least for spacings that are $\pm$50\% of the
initial value. This means that we must have $\lmin/W > 32$
because of the low-spacing limitation.

Next, $\lmin$ should not be much larger than this value,
since the computer time necessary to simulate the
evolution of an array of initial spacing $\lmin$ can 
be estimated to scale as $\lmin^5$ (number of grid 
points: $\lmin\times\lmin$; time for the interface 
to advance by one spacing: $\lmin/v_p$; using
$\lmin^2v_p = $const., we get 
$t_{CPU} \sim \lmin^5\sim {v_p}^{-5/2}$).
From Fig.~\ref{figjhunt}, we can see that $n=1$ and
$v_p=0.01$ give $\lmin$ of the right order of magnitude;
however, since $\lmax/W\sim 58$, the available range
of lamellar spacings is somewhat small (for an initial
spacing of $1.5\,\lmin$, nucleation would set in for
an increase of the local spacing by only 30\%).
Since the range of available spacings increases
with decreasing pulling speed, one possible solution
would be to further reduce $v_p$. However, as discussed
above the necessary computer time rapidly becomes prohibitive.

Another way out is to change the exponent in the equations 
for the mobilities, Eqs. (\ref{mobilities}).
If we choose $n>1$, the diffusivity
is increased in the whole interfacial region, whereas
it remains zero in the solid. This leads to higher diffusion
currents along the surface than for $n=1$. Hence, the
pileup of the rejected atoms at the interface is lower, and 
consequently $\lmax$ is higher. The price to pay is that this 
model, in its sharp-interface limit, is not equivalent to the
classical JH model, but contains additional surface 
diffusion terms \cite{Almgren99}. However, as shown in
Fig. \ref{figjhunt}, the qualitative behavior of the
undercooling versus spacing curve does not change.
For $v_p=0.01$, $\lmin$ is larger than the theoretical 
JH-value by about 10\%, whereas $\dtmin$ is about
15\% too low. On the other hand, $\lmax/W = 74$,
such that we now have at our disposal a sufficient
range of spacings. 

For these parameters, we observe for large spacings the 
well-known period-preserving 
oscillatory instability that sets in at about 2$\lmin$
\cite{Karma96}. Even beyond the threshold of this instability,
steady states can be reached to within an excellent 
precision, because we start from an exactly symmetric
initial condition and because the numerical noise of the
phase-field approach is extremely low. To trigger the
instability within a reasonable simulation time, 
an explicit perturbation that breaks the
symmetry between the two phases had to be added.
Such unstable steady states are shown as filled symbols 
in Fig.~\ref{figjhunt}.

The mechanism for lamella creation by nucleation is in fact very 
useful for the simulations of well-developed colonies where 
lamellae are frequently created at the solidification front.
We want to confront our simulations to the experimental
findings of Akamatsu and Faivre \cite{Akamatsu00}, who work 
with thin samples of a transparent eutectic alloy enclosed 
between parallel glass plates. In their experiments,
creation of new lamellae takes indeed place predominantly
in the center of already existing lamellae. However, the
detailed mechanism is still unknown. New lamella do not 
form by nucleation, since the interfacial undercoolings
are not high enough. Most likely, the ``pockets'' in the
center of large lamellae are ``invaded'' from pre-existing
neigboring lamellae of the opposite solid by fingers that
grow in the meniscus between the glass plates and the
growing solid. The point here is that the modeling of
such a process is out of reach for our present
computational resources, since it is necessarily
three-dimensional. Within the framework of two-dimensional
simulations, we simply need a criterion to decide when
new lamella should form, and the ``automatic'' implementation
of a maximal lamellar width $\lmax$ in our model is an 
adequate solution that avoids the implementation of an
explicit nucleation rule, as done for example in Ref. \cite{Lo01}.

\section{Linear stability of lamellar arrays}
\subsection{Theory}

We have recently performed a detailed linear stability analysis
of a lamellar eutectic interface in the presence of ternary impurities.
Rather than to repeat the calculations here, we will give a
brief summary of the main assumptions and results before
discussing the phase-field simulations. Our analysis is
an extension of the method used by Datye and Langer (DL) to
analyze the stability of lamellar arrays without impurities \cite{Datye81}.
It is based on a perturbation scheme for the Jackson-Hunt
solution and proceeds as follows.
\begin{enumerate}
\item The coordinates of the trijunction points are chosen as
fundamental variables to describe the state of the perturbed
system. This amounts to a ``discretization'' of the original
continuous system. Each trijunction point has two degrees
of freedom, namely its $x$ and $z$ positions (motion parallel
and normal to the isotherms, respectively).
\item For a lamellar interface that is gently curved on a
scale much larger than $\lambda$, the lamellae are assumed
to grow perpendicular to the envelope of the composite front 
(Cahn's hypothesis). This connects the time derivative of the local 
lamellar spacing to the shape of the front. For example, in a 
protrusion where the front curves outward, the local spacing 
increases during further growth.
\item Given the positions of the trijunction points, the actual
interface shape is replaced by a piecewise planar interface,
and a perturbed diffusion field is calculated. The Gibbs-Thomson
equation is then used to obtain an eigenvalue problem for normal
modes, i.e. perturbations proportional to $\exp(ikx+\omega t)$,
where $k$ is the wave vector of the periodic perturbation, and
$\omega$ its growth rate. The solutions of the eigenvalue equation
give the dispersion relations $\omega(k)$. Since there are four
degrees of freedom per lamella pair (two for each trijunction),
$\omega(k)$ has four branches. Of those, there are two that are
relevant for the long-wavelength instability we are interested in.
\end{enumerate}

It turns out that the final result of this rather complicated
analysis can be understood in terms of an effective front approach.
Namely, one can separate two scales: the local lamellar spacing, 
and the large-scale smooth envelope of the lamellar front. The
evolution of the local spacing is slaved to the shape of the
front by the assumption of normal motion (Cahn's
hypothesis). On the scale of the smooth front, the lamellar
structure introduces an interfacial undercooling that is
approximately given by the JH law, taken with the local spacing 
and interface velocity. Using these ingredients, it is possible
to include the lamellar geometry in the usual Mullins-Sekerka
stability analysis and to obtain the dispersion relation. This
result can be recovered from the more complicated discrete
analysis, with one additional ingredient. The eutectic diffusion
field that governs the exchange of atoms between neighboring
lamellae gives, when perturbed on a large scale, a stabilizing
contribution to the total interfacial undercooling. The functional
form of this stabilization is the same as for the surface tension
terms, and this effect can therefore be included in the simple
effective front approach by simply ``renormalizing'' the capillary
length.

The two main results of this analysis are that (i) the instability
threshold is close to the well-known constitutional supercooling
criterion, with a small capillary correction, and (ii) in contrast
to the Mullins-Sekerka instability, where unstable modes always
have real growth rates, the lamellar structure may lead to
complex growth rates, and hence to oscillatory modes.
The origin of these oscillations can be understood as follows:
in a protrusion of the front, the lamellar spacing increases. 
This leads to a local change in the JH undercooling 
that, for a small distortion of an array of spacing $\lambda_0$,
is proportional to the slope of the JH plot. 
For $\lambda_0 > \lmin$, this provides
a ``restoring force'' for the large-scale front. Since only
the {\em change with time} of the lamellar spacing (but not
the spacing itself) is related to the shape of the front,
the dispersion relation becomes quadratic in $\omega$,
instead of the linear Mullins-Sekerka equation. 
There are two solutions to this equation for each wave 
vector $k$. In physical terms, this is the consequence 
of the additional ``internal degree of freedom'' $\lambda$ 
of the front. As discussed in detail in Ref. \cite{Plapp99}, 
real and complex growth rates may occur, depending on 
the ratio $G/v_p$, the lamellar spacing, and the impurity
content. For large enough spacings,
when the ``restoring force'' mentioned above is strong enough,
we expect that the complete dispersion relation is complex. One of
the goals of the present paper is to test this prediction
by direct simulation of the basic equations of motion.

\subsection{Single mode simulations}
Let us first study the behavior of a single unstable mode
of a lamellar array with initial spacing $\lambda_0$.
The parameters besides $\lambda_0$ that control stability
are the impurity content and the ratio $G/v_p$. We define
the dimensionless parameters
\begeq
\Lambda=\lambda_0/\lmin,
\endeq
\begeq
w = {\mtil \Delta\ctil\over m_u\Delta u} = 
    {\mtil \ctilinf (1/K-1)\over m\Delta c},
\endeq
\begeq
g = {2 D G\over v_p m_u\Delta u} = {2 D G\over v_p m\Delta c}.
\endeq
Here, $\lmin$ is obtained from an interpolation of our
simulation data shown in Fig. \ref{figjhunt}. For $n=4$
and $v_p=0.01$, $\lmin\approx 34$. The freezing ranges
of the eutectic and the impurities are, expressed in the
parameters of our model, $m_u\Delta u = 2T_E/\alpha$, and
$\mtil \Delta \ctil = \mtil (1/K -1)\ctilinf = 
b(1-K)^2T_E\ctilinf/(K\alpha)$.

A lamellar array is prepared by replicating
the steady-state solution for one lamella pair $N$ times.
We apply a cosine perturbation 
to the steady state and impose the initial condition
\begeq
\phi(x,z,0) = \phi_0(x,z+A_0\cos(2\pi x/N)),
\label{perturb}
\endeq
where $\phi_0(x,z)$ is the steady-state solution. The other
fields ($u$ and $\ctil$) are perturbed in the same
manner. The perturbation amplitudes $A_0$ are usually
much smaller than the interface width (typically, $A_0/W<0.1$),
and the values on the grid points are obtained by linear
interpolation of the numerical steady-state solution.

To analyze the evolution of the system, we store periodically 
the positions of all the interfaces (solid-solid and solid-liquid).
In addition, we determine the positions of all the trijunction
points by searching the intersections of the level curves
$\phi=0$ and $u=0$. The coordinates of the trijunction point
to the left of the $\nu$-lamella ($\nu = \alpha,\beta$) in
the lamella pair number $i$ are labeled $(x_i^\nu,z_i^\nu)$.
We define the deviations of the trijunction
point coordinates from their steady-state values,
\beglett
\label{coord}
\begin{eqnarray}
\xi^\nu_i & = & z^\nu_i - \bar z \\
y^\nu_i & = & x^\nu_i - \bar x^\nu_i
\end{eqnarray}
\endlett
as well as their discrete Fourier transforms,
\begin{eqnarray}
X_\nu(\kappa,t) & = & {1\over N} \sum_{n=1}^N \xi^\nu_n \exp(2\pi i \kappa n) \\
Y_\nu(\kappa,t) & = & {1\over N} \sum_{n=1}^N y^\nu_n \exp(2\pi i \kappa n),
\end{eqnarray}
where $\kappa = k\lambda_0/(2\pi)$ is a dimensionless wave vector.

\begin{figure}
\centerline{
  \psfig{file=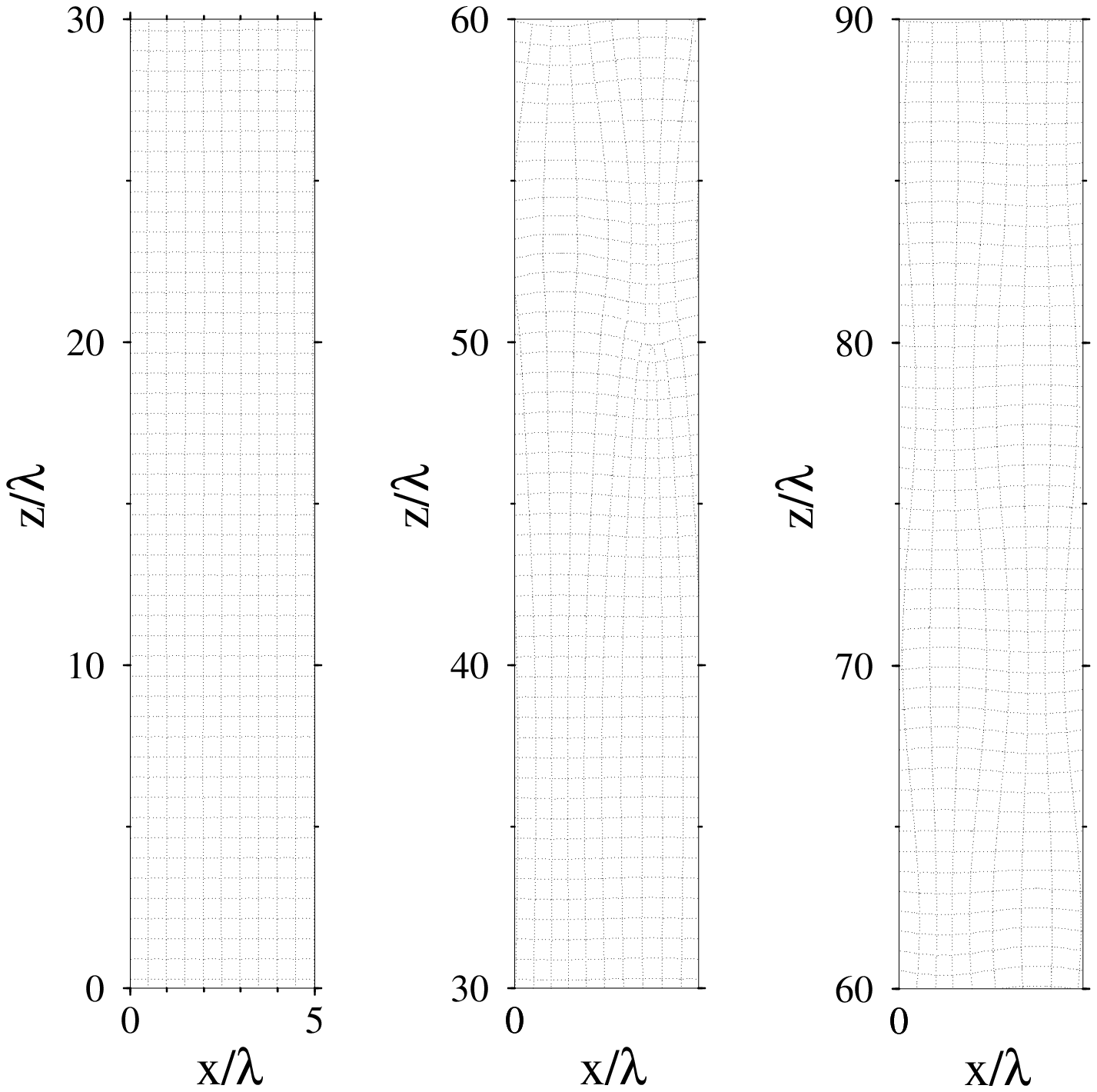,width=.45\textwidth}}
\bigskip
\caption{Evolution of oscillatory modes. Growth direction
is from bottom to top, and three successive frames are
shown from left to right. Shown are the solid-solid-interfaces,
as well as successive snapshot pictures of the solid-liquid interfaces.
The system is perturbed with a single cosine mode of dimensionless wave 
vector $\kappa=0.2$. Simulation parameters: $v_p=0.01$, $G=0.0005$, 
$\lambda_0=40$, $K=0.5$, $\ctilinf=0.08$, $b=10$, giving
$\Lambda=1.175$, $w=0.2$, $g=0.05$.}
\label{figosc}
\end{figure}
\begin{figure}
\centerline{
  \psfig{file=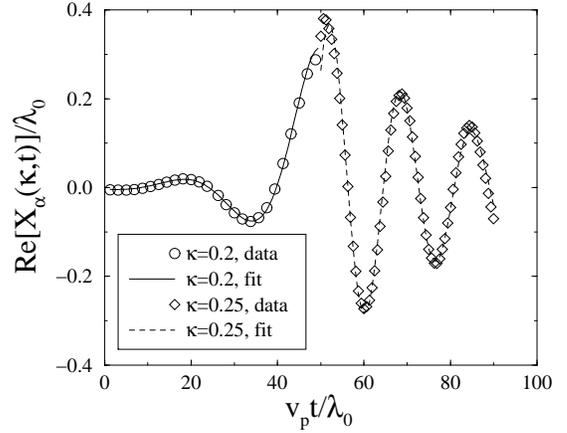,width=.4\textwidth}}
\medskip
\caption{Real part of the Fourier amplitude $X_\alpha(\kappa,t)$ as
a function of time, for the run of Fig. \ref{figosc}. Also shown
are the fits to growing (decaying) oscillating exponentials.}
\label{figamp}
\end{figure}
In Fig. \ref{figosc}, we show the evolution of an array of
five lamellae, started in a single mode with $\kappa=0.2$.
An oscillatory $5$-$\lambda$ mode develops.
Its amplitude grows until a lamella termination occurs
at $z/\lambda_0 = 50$. Subsequently, the system shows
a decaying $4$-$\lambda$-oscillation and approaches a
steady-state solution with 4 lamella pairs. To analyze
this evolution, we use the Fourier components $X_\nu(\kappa,t)$.
Since the initial perturbation is not proportional to the
(unknown) eigenvector corresponding to a single mode of
the complete (continuous) system, the Fourier components
for $\kappa\neq 0.2$ will not remain zero. However, they remain
sufficciently small to be neglected in the data analysis.
In Fig. \ref{figamp}, we show the evolution of 
${\rm Re}[X_\alpha(\kappa,t)]$ versus time, for $\kappa=0.2$
before the lamella elimination, and for $\kappa=0.25$ afterwards
(note that the elimination of one lamella pair corresponds
to a change in the unperturbed lamellar spacing $\lambda_0$).
Oscillating modes correspond to complex growth rates.
We define the dimensionless growth rate by
\begeq
\Omega= \omega\lambda_0/v_p = \Omega_r + i\Omega_i,
\endeq
with $\Omega_r$ and $\Omega_i$ real. The growth rate is 
determined by a fit of the data to the function
\begeq
{\rm Re}[X_\alpha(\kappa,t)] = A \exp(\Omega_r t) \sin[\Omega_i (t-t_0)],
\endeq
where $t$ is measured in units of $\lambda_0/v_p$. In practice,
we obtain a value of $t_0$, i.e. the time of one of the zero 
crossings, by numerical interpolation, and then use a least-squares
fitting procedure with $A$, $\Omega_r$, and $\Omega_i$ as free 
parameters. As can be seen from Fig. \ref{figamp}, the fit is
excellent. Surprisingly, the fit remains accurate up to the
immediate vicinity of the lamella termination event. This
indicates that the system is well described by a single,
exponentially growing mode even for large deformations of
the initial array. In particular, the linearization that is
the basis for the theoretical analysis remains
valid even if the lateral displacements are large, i.e., 
$y^\nu_i/\lambda_0 \sim 1$.

\subsection{Dispersion relations}
\begin{figure}
\centerline{
  \psfig{file=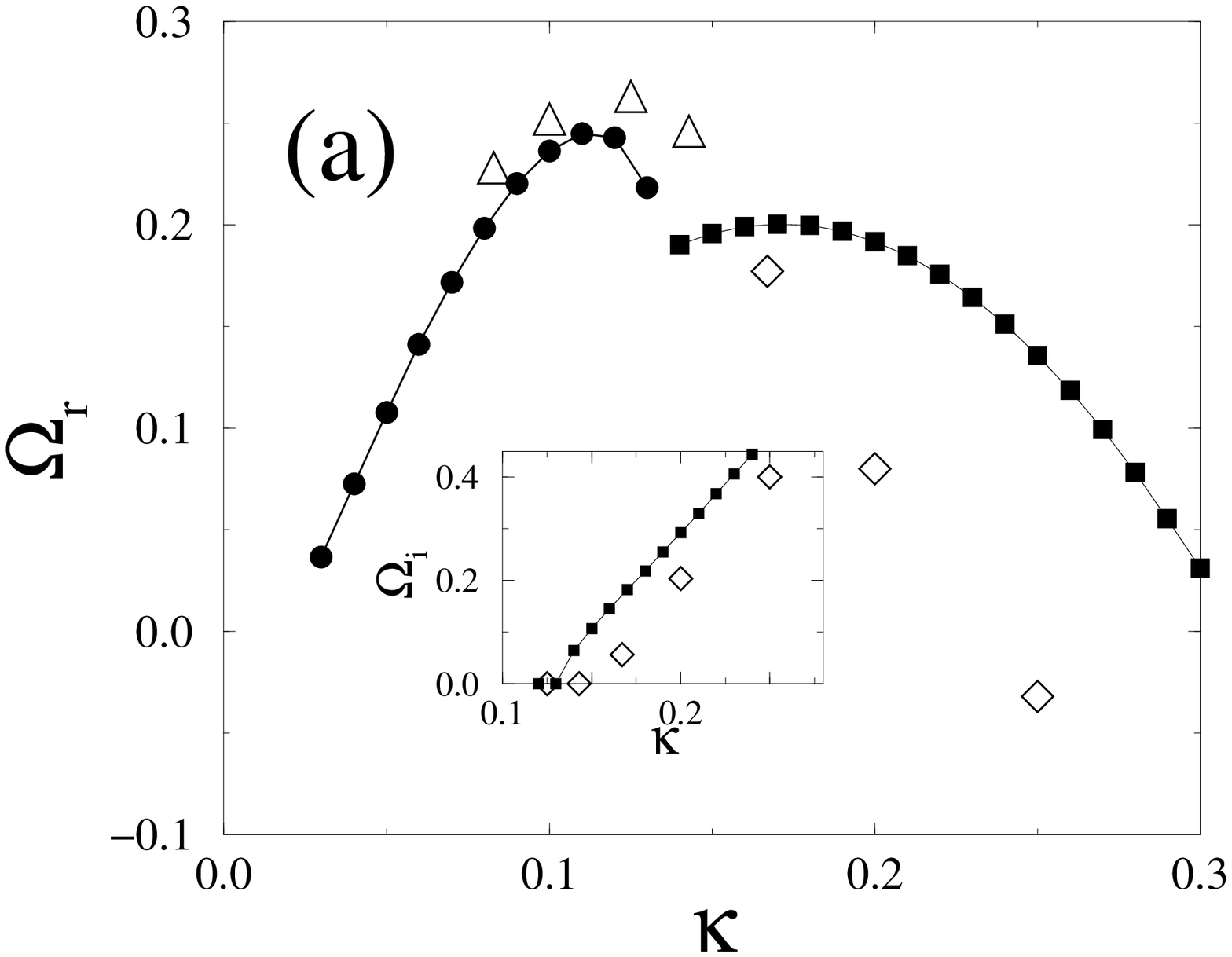,width=.45\textwidth}}
\smallskip
\centerline{
  \psfig{file=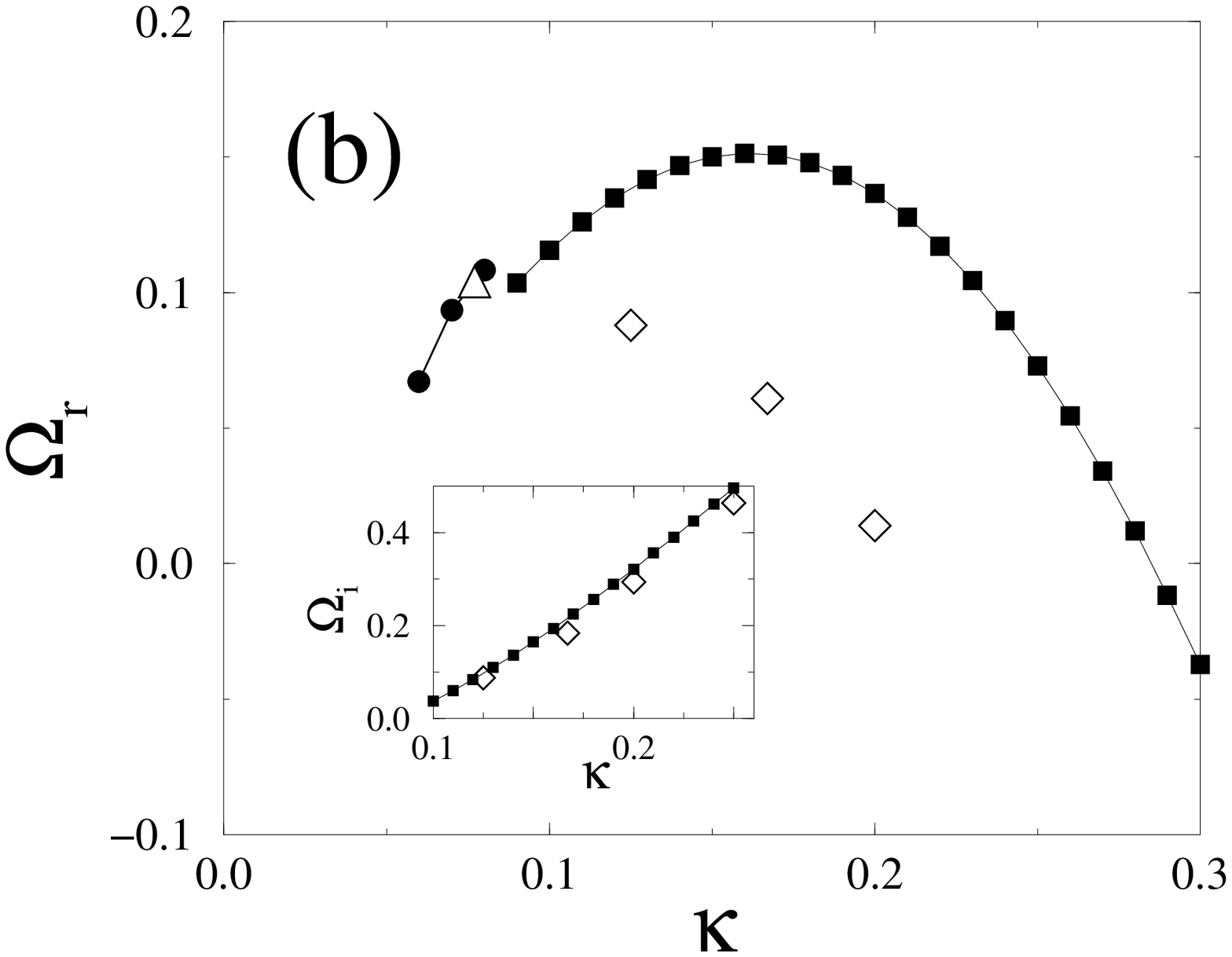,width=.45\textwidth}}
\medskip
\caption{Plots of the dimensionless growth rate 
$\Omega$ versus dimensionless wave vector $\kappa$.
The main graphs show the real part $\Omega_r$, 
the insets the imaginary part $\Omega_i$
for (a) $\Lambda=1.175$, $w=0.2$, $g=0.05$
and (b) $\Lambda=1.175$, $w=0.2$, $g=0.1$. Filled symbols and
lines: theoretical predictions from Ref.~\protect\cite{Plapp99};
circles: real modes ($\Omega_i=0$), squares: complex modes
($\Omega_i\neq 0$). Open symbols: simulation data; triangles:
real modes, diamonds: complex modes.
}
\label{figdisper}
\end{figure}

The simulation and fitting procedures outlined above were
carried out for various values of the control parameters 
and arrays of different sizes to construct the dispersion
relations $\Omega(\kappa)$. In Fig. \ref{figdisper}, we
show a comparison of the obtained data with the theoretical
predictions of Ref. \cite{Plapp99} for two different values
of the temperature gradient. For both dispersion relations,
there are stationary ($\Omega$ real) and oscillatory ($\Omega$
complex) modes. According to theory, for $g=0.05$ the fastest 
growing mode is stationary, whereas for $g=0.1$ is is oscillatory.

In all cases, the nature of the mode (stationary or oscillatory)
agrees with the theoretical predictions. Furthermore, the
oscillation frequency of the complex modes ($\Omega_i$)
is always in good quantitative agreement with theory.
In contrast, the growth rates ($\Omega_r$) are in good
agreement only for small wave numbers; for large wave
numbers, the simulated growth rates
are systematically much smaller than predicted by theory,
and the difference increases with the dimensionless wave vector.
Therefore, in the simulations at $g=0.1$, the fastest growing
mode is stationary, and not complex as predicted by theory. 
For $\Lambda=1.47$ and $g=0.1$, we obtain a stability spectrum 
that is entirely complex (data not shown),
both in theory and simulations.

Just as the JH theory, our stability analysis of a lamellar
array contains several simplifying assumptions. It is therefore
necessary to check whether the differences between theory and
simulations are due to the approximations made in the stability
analysis, or due to the phase-field approach, which is a genuine
representation of the original free-boundary problem only in
the limit $W/\lambda \to 0$. Therefore, we focused on a single
complex mode at $g=0.1$ and $\kappa=0.2$ (5-$\lambda$-oscillation)
and conducted a series of runs with decreasing pulling speed $v_p$.
Since $\lmin\sim v_p^{-1/2}$, we increased the spacing $\lambda_0$
to keep the reduced spacing $\Lambda$ constant. The temperature
gradient $G$ was also decreased to keep $g$ constant. 

\begin{figure}
\centerline{
  \psfig{file=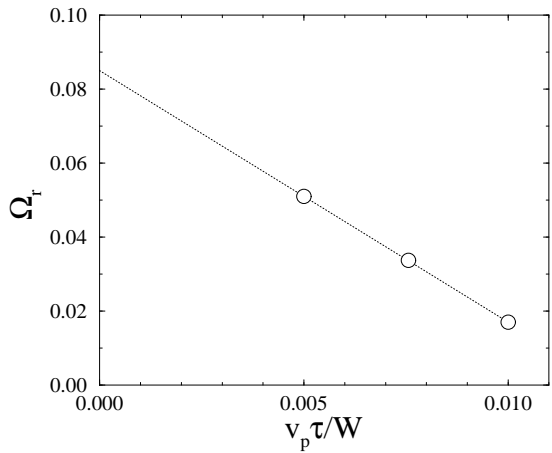,width=.4\textwidth}}
\medskip
\caption{Real part of the dimensionless growth rate 
$\Omega$ versus pulling speed for $\Lambda=1.175$,
$g=0.1$, $w=0.2$, and $\kappa=0.2$. Symbols: simulation
result. Dashed line: linear extrapolation to $v_p=0$.
}
\label{figextra}
\end{figure}

The results for the growth rate $\Omega_r$ versus $v_p$ are shown in 
Fig.~\ref{figextra}. The data fall on a straight line, and by
extrapolation to $v_p=0$ we find $\Omega_r(v_p=0) = 0.085$.
In contrast, the variation of the oscillation frequency is
very small (from $\Omega_i = 0.291$ at $v_p=0.01$ to
$\Omega_i = 0.302$ at $v_p=0.005$).
This linear variation of $\Omega_r$ with $v_p$ 
indicates that the dominant corrections to the
sharp-interface limit of the phase-field model
scale as $W/l_D=Wv_p/D$. Corrections in the other
involved small ratios, $W/\lambda_0$ and $\lambda_0/l_D$,
seem to be subdominant, since both scale as 
$1/\sqrt{v_p}$ at constant $\Lambda$.
An example for a correction that scales as $W/l_D$ is 
the interface kinetics; however,
inserting a kinetic term in the Mullins-Sekerka analysis
does not lead to a linear variation of the growth rate
with the kinetic coefficient. The solute trapping effect
also scales as $W/l_D$, but since it is quite involved
to evaluate its influence on the growth rates,
we have not investigated this issue in more detail.
We checked, however, that the variation of
$\Omega_r$ with $v_p$ is not a consequence of the surface
diffusion term introduced by our choice $n=4$ in the
mobility function: a simulation with $n=1$ and comparable
$\Lambda$ yielded similar results.

\begin{figure}
\centerline{
  \psfig{file=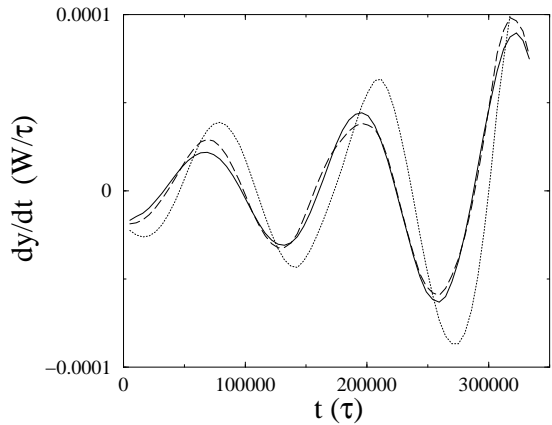,width=.4\textwidth}}
\medskip
\caption{Sideways velocity, $dy/dt$, of one trijunction
point for $\Lambda=1.175$, $g=0.1$, $w=0.2$, $\kappa=0.2$,
and $v_p=0.00751$. Solid line: data extracted from the
simulated curve $y(t)$. Dotted line: prediction of
Cahn's hypothesis. Dashed line: best fit to 
Eq.~(\protect\ref{cahnfit}).
}
\label{figcahn}
\end{figure}

The simulated growth rate, extrapolated to $v_p=0$,
is still markedly different from the theoretical prediction
$\Omega_r = 0.1365$. We therefore checked several assumptions
that are used in the linear stability analysis, in particular
Cahn's hypothesis that the lamellae always grow perpendicular
to the large-scale front. Expressed in terms of the
trijuction point coordinates defined in Eq. (\ref{coord}), 
this assumption reads
\begeq
\partial_t y^\beta_n = -{v_p\over\lambda_0}
                   \left(\xi^\alpha_{n+1}-\xi^\alpha_n\right)
\label{cahn}
\endeq
for the trijunction point to the left of the $n$th $\beta$ 
lamella. From the simulation data, explicit values
of $\partial_t y^\beta_n$ and the vertical displacements
$\xi_n^\alpha(t)$ are available, and Eq.~(\ref{cahn})
can be directly checked. As shown in Fig.~\ref{figcahn},
Cahn's hypothesis is clearly violated. We tried to fit
the difference of the right-hand-side and left-hand-side
of Eq. (\ref{cahn}) to various functions of the trijunction
coordinates and found that the modified equation
\begeq
\partial_t y^\beta_n = -{v_p\over\lambda_0}
                   \left(\xi^\alpha_{n+1}-\xi^\alpha_n\right) 
                       -{v_p\over\lambda_0} B y^\beta_n
\label{cahnfit}
\endeq
yields a good fit with a single adjustable parameter $B$,
as is shown in Fig.~\ref{figcahn}.

\begin{figure}
\centerline{
  \psfig{file=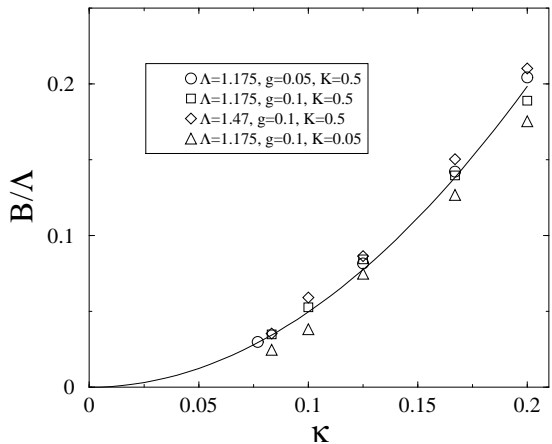,width=.4\textwidth}}
\medskip
\caption{Fit parameter $B$ divided by the reduced spacing
$\Lambda$ versus wave vector $\kappa$ for various sets
of control parameters. The line is a best fit of all
data with Eq.~\protect\ref{quadratic}.
}
\label{figb}
\end{figure}

We repeated the above fit for all our simulation data.
Rather remarkably, the correction given by Eq. (\ref{cahnfit}) 
works both for oscillatory and stationary modes, and
the fit parameter $B$ behaves smoothly in the crossover
regions. This shows that the violation of Cahn's hypothesis
is a consequence of the local front geometry, and not
a cooperative effect depending on the nature of the mode.
We found that the fit parameter $B$ mainly depends on the 
wave vector $\kappa$ and the reduced spacing $\Lambda$.
We also found weak dependencies on the temperature gradient
$g$ and the impurity partition coefficient $K$.
More importantly, $B$ is almost independent
of the ratios $\lambda_0/W$ and $W/l_D$: for the series of 
runs of Fig.~\ref{figextra}, $B$ varied between $0.240$ for
$v_p=0.01$ and $0.215$ for $v_p=0.005$. Even though
$B$ decreases with $v_p$, it does not extrapolate to $0$,
such that this effect is not an artefact of the phase-field
model. In Fig.~\ref{figb} we show the fitted values of $B$, 
rescaled by the reduced spacing $\Lambda$, versus the
wave number $\kappa$ for various external parameters.
Neglecting the weak dependencies on $K$ and $g$,
a fit to all the data points yields
\begeq
B = B_0 \Lambda \kappa^2,
\label{quadratic}
\endeq
with $B_0 = 4.96$. 

Although Cahn's rule is violated, the resulting deviations 
of the growth angles from $90^\circ$ are very small. 
To see this, let us use the geometrical relation 
$\partial_t y^\beta_n = -v_p \tan \delta^\beta_n$,
where $\delta^\beta_n$ is the angle between the
solid-solid interface at the trijunction and the
$z$ direction. From Eq. (\ref{cahnfit}) we can
calculate the deviation of $\delta^\beta_n$ from
the value predicted by Cahn's rule, which is
$(\xi^\alpha_{n+1}-\xi^\alpha_n)/\lambda_0$.
In our simulations, this deviation never exceeded $1^\circ$. 
Due to the finite interface width of the phase-field model, 
it is very difficult to measure angles directly at
the trijunction points, and such small deviations
cannot be resolved. Therefore, the procedure outlined
above that uses the whole trajectory of a trijunction
point is the only way to obtain quantitative 
information about the violation of Cahn's rule
directly from the simulations. It should
be emphasized, however, that while the deviation
itself is small, since the growth angle $\delta^\beta_n$
is itself small, the {\em ratio} of the two is not
necessarily small. Indeed, it can be seen from
Fig. \ref{figcahn} that the correction constitutes
a sizeable fraction of the growth angle. This
explains why such a small deviation can induce
quite large shifts in the stability spectrum.

The functional form of Eq.~(\ref{quadratic}) allows us
to draw several interesting conclusions. First, since
the coefficient $B$ is proportional to the reduced
eutectic spacing $\Lambda$, but almost independent
of the impurity content, this effect is not specific
to ternary alloys, but should also occur in binary
eutectics. Secondly, a deviation from Cahn's rule
has been previously reported in binary eutectics
\cite{Caroli90,Chen01}. However, this deviation
becomes important only for very
strong temperature gradients, i.e. for $g\gg 1$,
when the front is almost flat, except in the immediate
vicinity of the trijunction points. Here, we are 
in the opposite regime, with $g\ll 1$, and a front that 
consists of round arcs between the trijunction points;
since, in addition, $B$ is almost independent of $G$,
we conclude that the effect described by
Eqs.~(\ref{cahnfit}) and (\ref{quadratic}) seems not
to be captured by these calculations.

Additional physical insight can be gained by inserting back
Eq.~(\ref{quadratic}) into Eq. (\ref{cahnfit}). The correction 
to Cahn's rule is proportional to $\kappa^2 y$. In a
continuum limit, where the function $y(x)$ is a smooth
interpolation of the lateral trijunction displacements,
this corresponds to a second derivative, $\partial_{xx} y(x)$.
For $y$ varying slowly on the scale of $\lambda_0$, we
have $\lambda(x)\sim\lambda_0(1+\partial_x y)$, such that
the correction to Cahn's rule is proportional 
to the gradient of the local spacing. 
The motion of the trijunctions is therefore
a combination of the perpendicular lamellar growth
considered before and a small lateral drift proportional 
to the gradient of the local spacing. While such a term
certainly appears to be reasonable, a more detailed
theoretical analysis is clearly warranted.

This violation of Cahn's hypothesis explains the remaining
discrepancies between our simulation results and the theory.
To modify the theory by the inclusion of the corrective
term in Eq. (\ref{cahnfit}) seems possible, but is
out of the scope of the present paper.

\section{Dynamics of colony formation}
To study the instabilities that lead to the formation
of colonies, we constructed large arrays as described
before, and perturbed the steady-state solution by
a spatial displacement of the fields along the
$z$ direction. The amplitude of the displacement
was a random variable of $x$ with a white noise
spectrum and an amplitude comparable to one lattice
spacing. The goal was to study the initial instability
of such random arrays as well as the nonlinear dynamics 
of well-developed colonies.
The latter required long runs in big systems. The
necessary computational power was attained by porting
our simulation code on a parallel CRAY T3E computer.
We used a simple domain-decomposition scheme for
parallelization, i.e. every processor calculated
a part of the system. A load-balancing algorithm 
that adjusted the domain boundaries as a function
of the computational load for each processor was
used to optimize the yield.

\begin{figure}
\centerline{
  \psfig{file=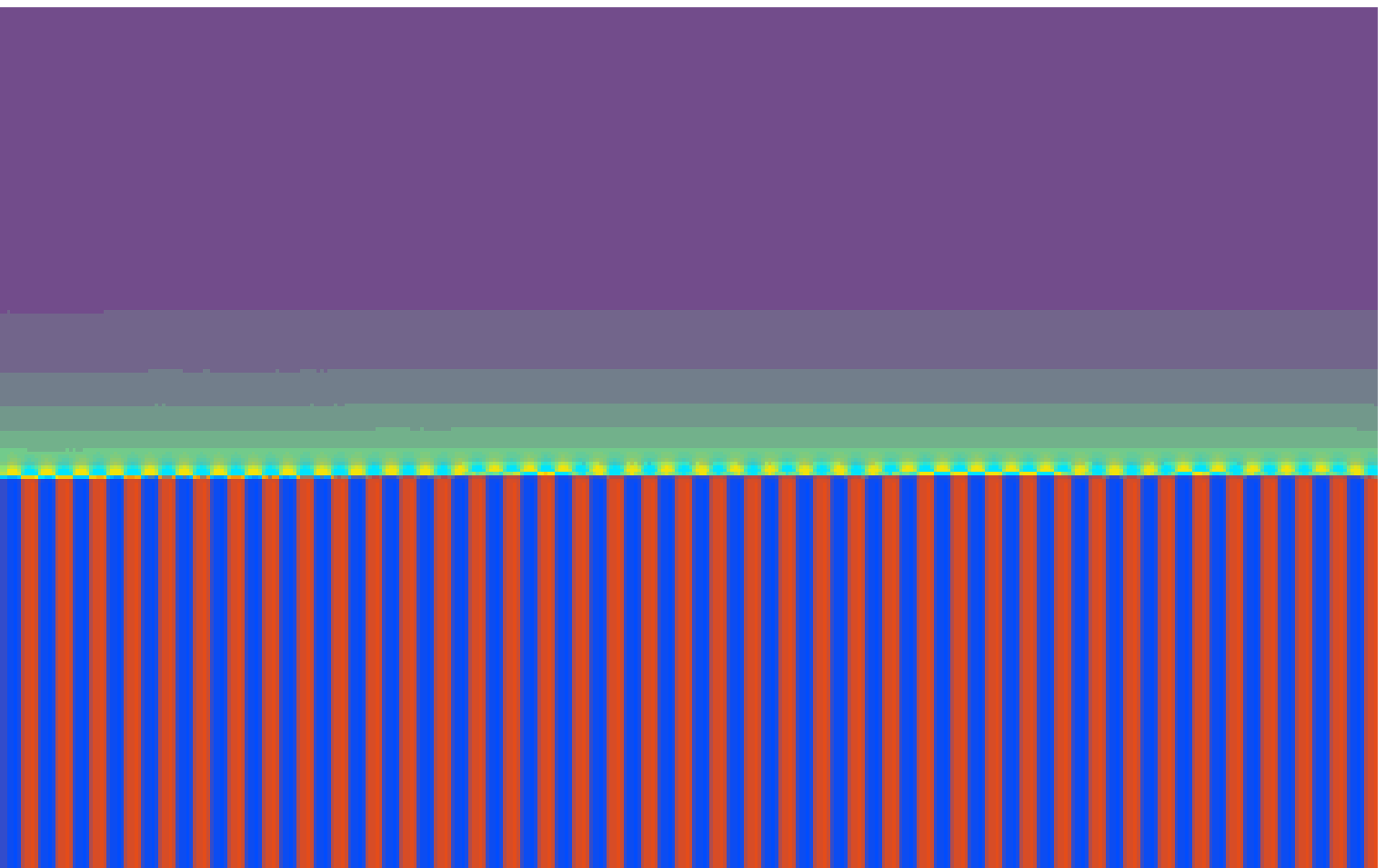,width=.45\textwidth}}
\medskip
\centerline{
  \psfig{file=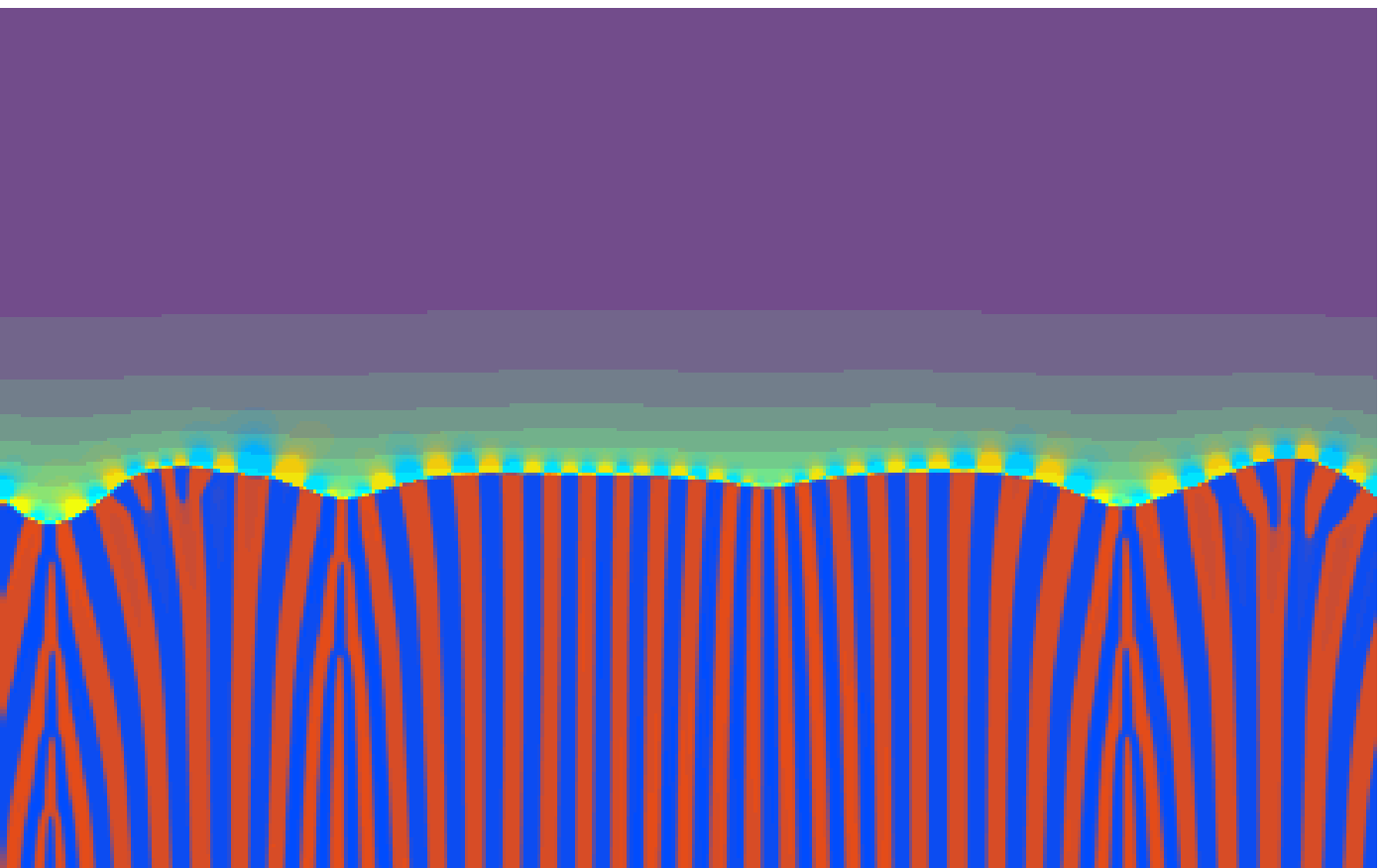,width=.45\textwidth}}
\medskip
\centerline{
  \psfig{file=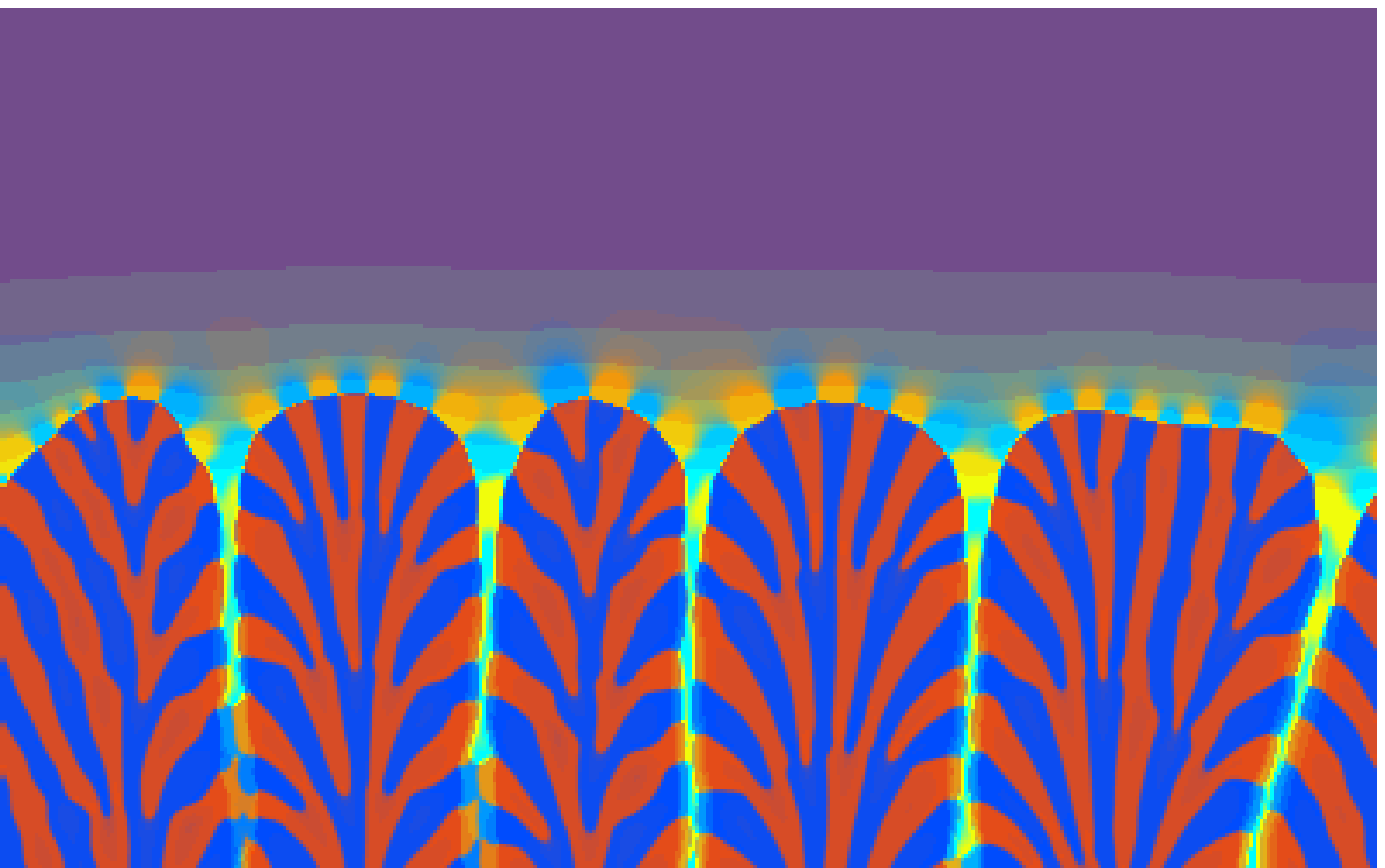,width=.45\textwidth}}
\medskip
\caption{Snapshot pictures of a run with 40 lamella pairs at 
eutectic composition and $g=0.1$, $\Lambda=1.175$, and $w=0.2$.
From top to bottom: $t/\tau=0, 100000, 375000$.
In the solid, red and blue represent the 
two solid phases. In the liquid, the green intensity is 
proportional to the impurity concentration; the small blue 
and yellow ``halos'' in advance of the growing lamellae are 
a visualization of the interlamellar (eutectic) diffusion
field.}
\label{figeut}
\end{figure}

\begin{figure}
\centerline{
  \psfig{file=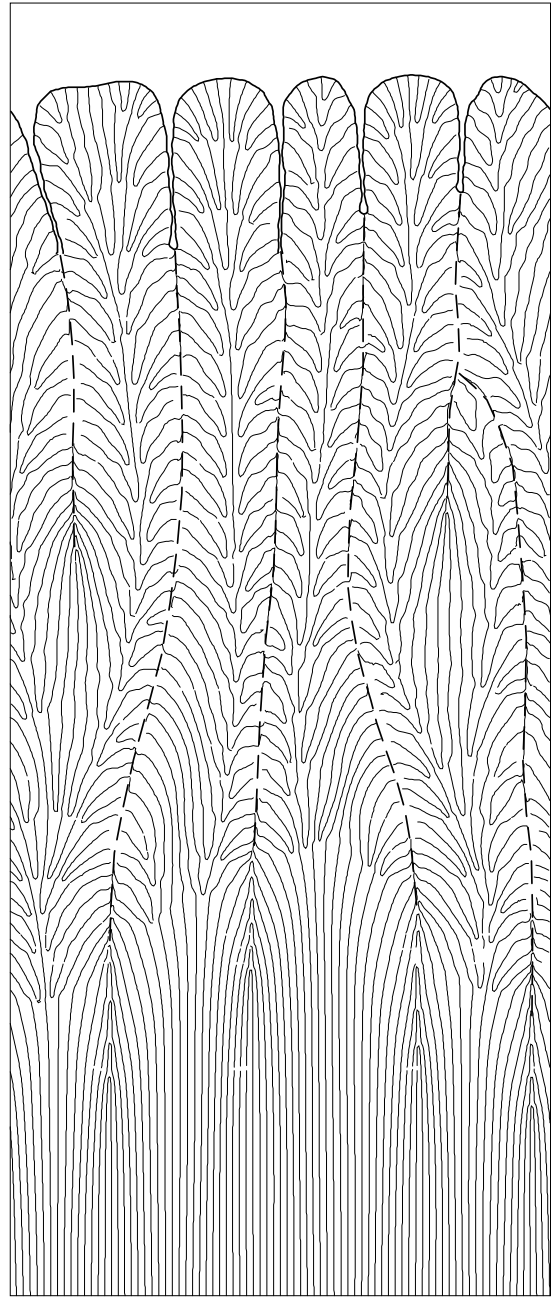,width=.4\textwidth}}
\medskip
\caption{Global view of the same run as in Fig. \protect\ref{figeut},
without greyscale. Thin lines: solid-solid interfaces. Thick solid
line: final solid-liquid interface. Thick dashed lines: trajectories
of the grooves between fingers. There are two tip-splitting and one 
finger overgrowth event. Note the concave part of the final front
in the center of the leftmost finger: a tip-splitting event will
soon take place.}
\label{figeutspacetime}
\end{figure}

For $u_\infty=0$ (eutectic composition), the initial 
evolution of the lamellar array is a linear superposition 
of the long-wavelength modes desribed in the
previous section. That is, if we decompose the set of
trijunction displacements into Fourier modes, each mode
grows with the (real or complex) growth rate that was
determined in the single-mode simulations of the preceding
section. In Fig. \ref{figeut}, we show the resulting
evolution for the same control parameters as in 
Fig. \ref{figdisper}(a). The fastest growing mode is
real with a wavelength of about $12\lambda_0$.
Indeed, this mode dominates the interface shape in the 
second snapshot, where the first lamella termination events 
have occured. At later times, the linear description becomes
evidently invalid. The further evolution is characterized
by the growth of long protruding fingers, as can be seen 
in the last snapshot picture. These fingers, however, do not
reach a steady-state configuration up to the end of our
simulation: their shape continuously changes, and there are
some tip-splitting and overgrowth events. To highlight this
feature, we show in Fig. \ref{figeutspacetime} a complete plot
of the whole solidified sample, where we have omitted the
greyscale for clarity, and where we have marked the trajectories
of the ``deep grooves'' between neighboring fingers. This
run was performed on a lattice of size $1600\times 1200$
and totals $15\times 10^6$ iterations. On the CRAY T3E,
this run required about 3000 hours single processor CPU time.

\begin{figure}
\centerline{
  \psfig{file=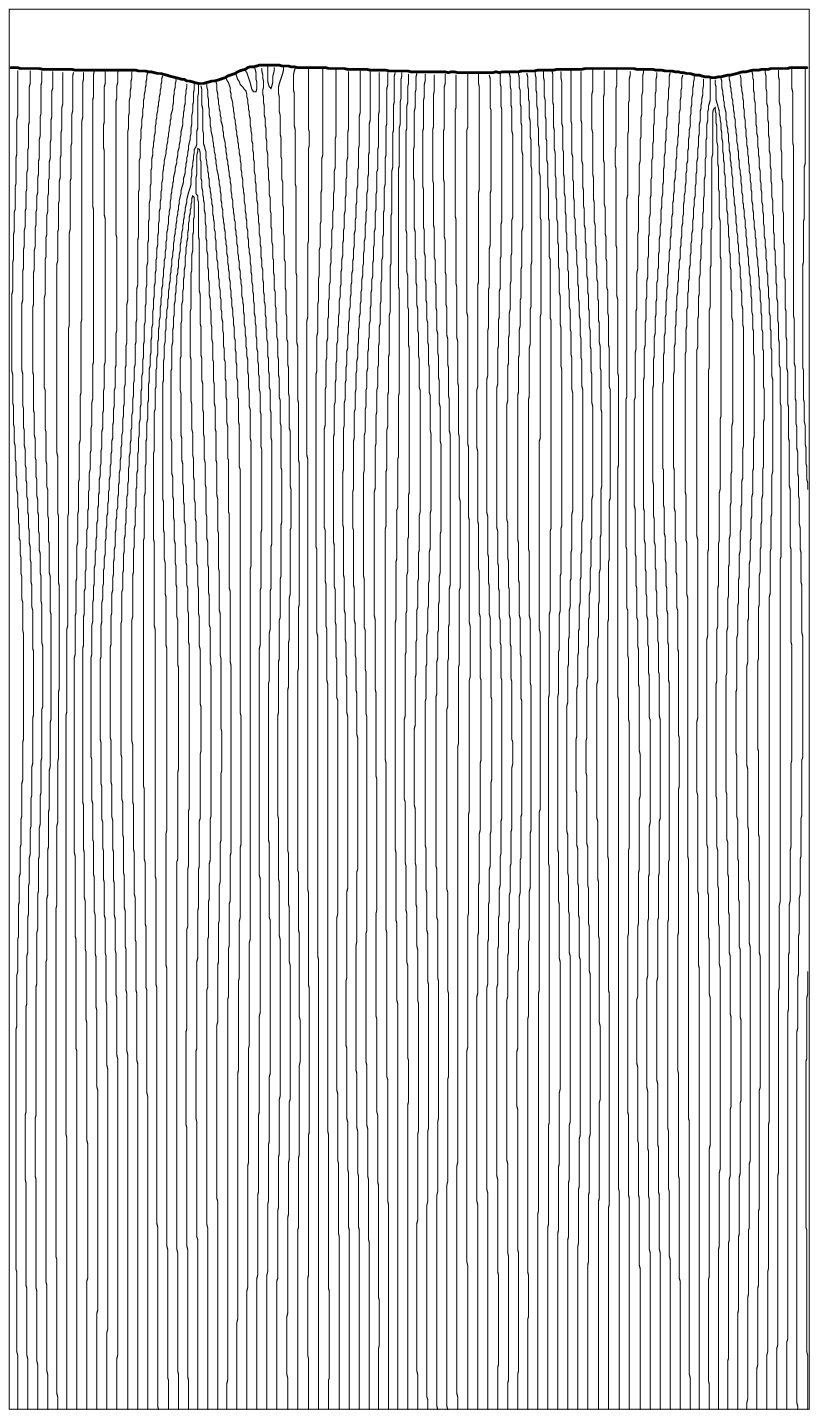,width=.4\textwidth}}
\medskip
\caption{Run with 40 lamella pairs at eutectic composition and 
$g=0.2$, $\Lambda=1.5$, and $w=0.2$. The dispersion relation is
entirely complex, and oscillatory patterns appear. Thin lines: 
solid-solid interfaces. Thick solid line: final solid-liquid 
interface.}
\label{figeut2}
\end{figure}

In Fig.~\ref{figeut2}, we show a run with again 40 lamellae,
but now with both a larger temperature gradient and a larger
initial spacing. Under these conditions, the instability
develops more slowly, and the dispersion relation is
entirely complex, such that we expect propagating or
oscillatory modes. Indeed, on the right side of Fig. \ref{figeut2},
there is an oscillatory ``breathing mode'' with wavelength
about $10\lambda_0$, whereas on the left side, a travelling 
perturbation of the lamellar pattern can be seen. The run
was not continued after the first lamella termination events,
since the nonlinear regime is expected to lead to similar 
fingered patterns as in Fig. \ref{figeut}.

\begin{figure}
\centerline{
  \psfig{file=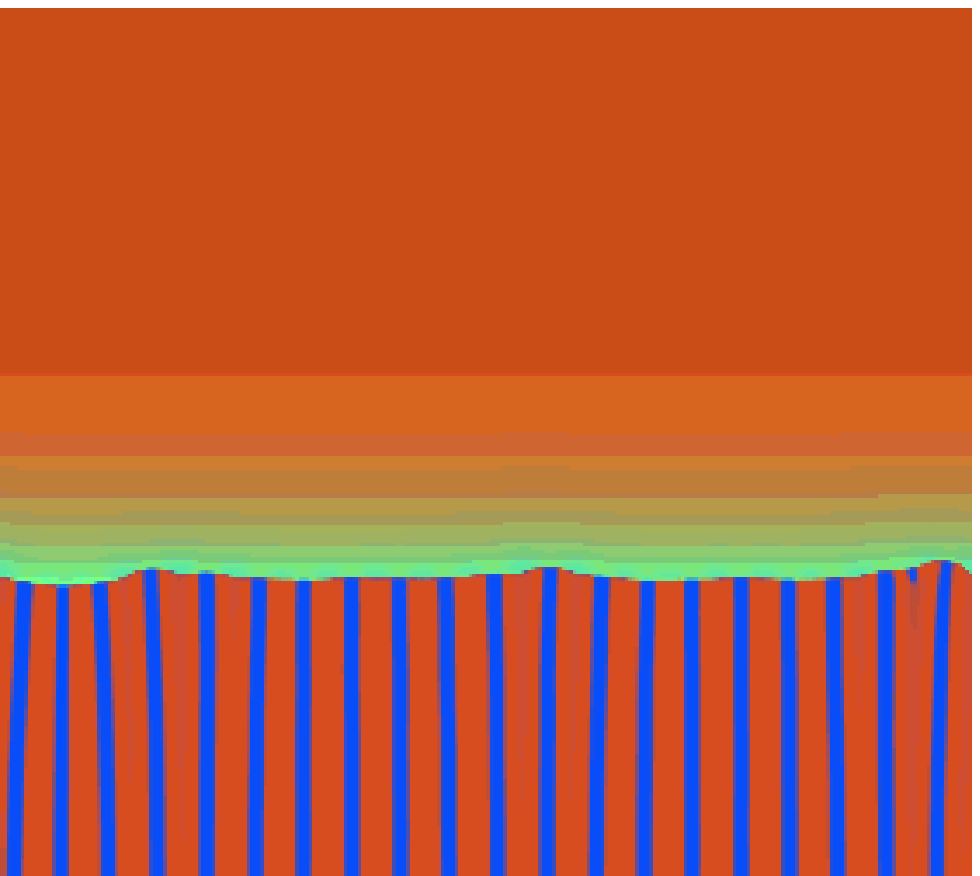,width=.32\textwidth}}
\medskip
\centerline{
  \psfig{file=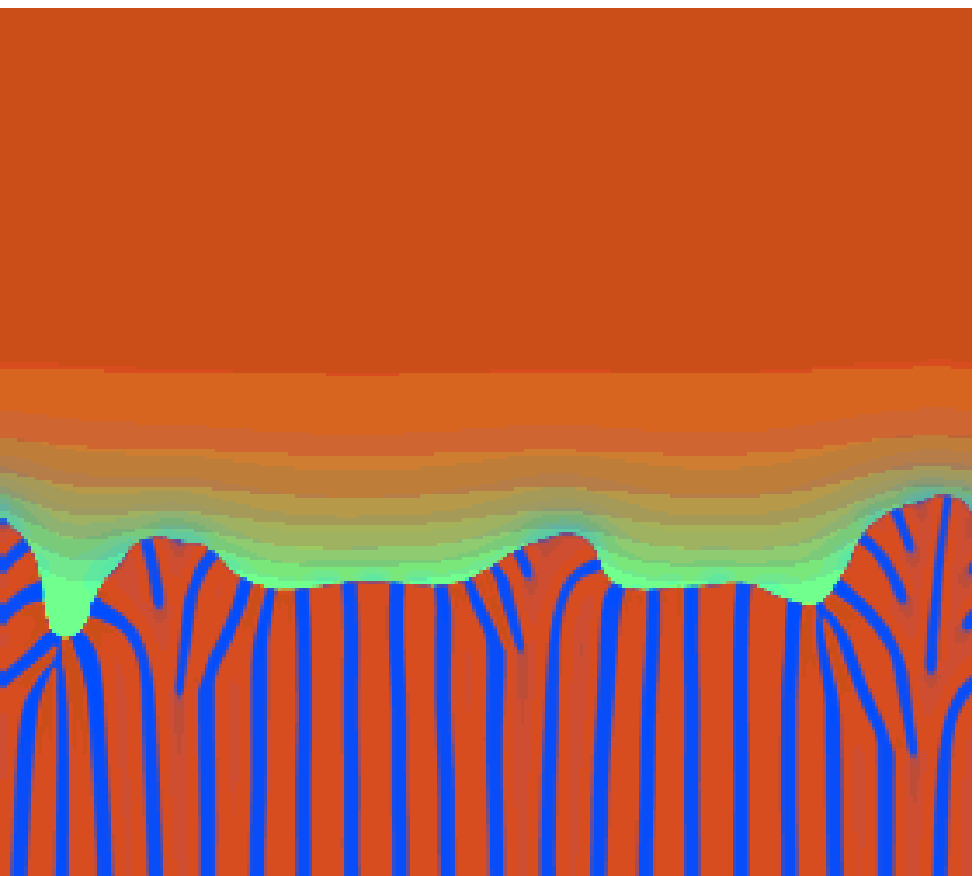,width=.32\textwidth}}
\medskip
\centerline{
  \psfig{file=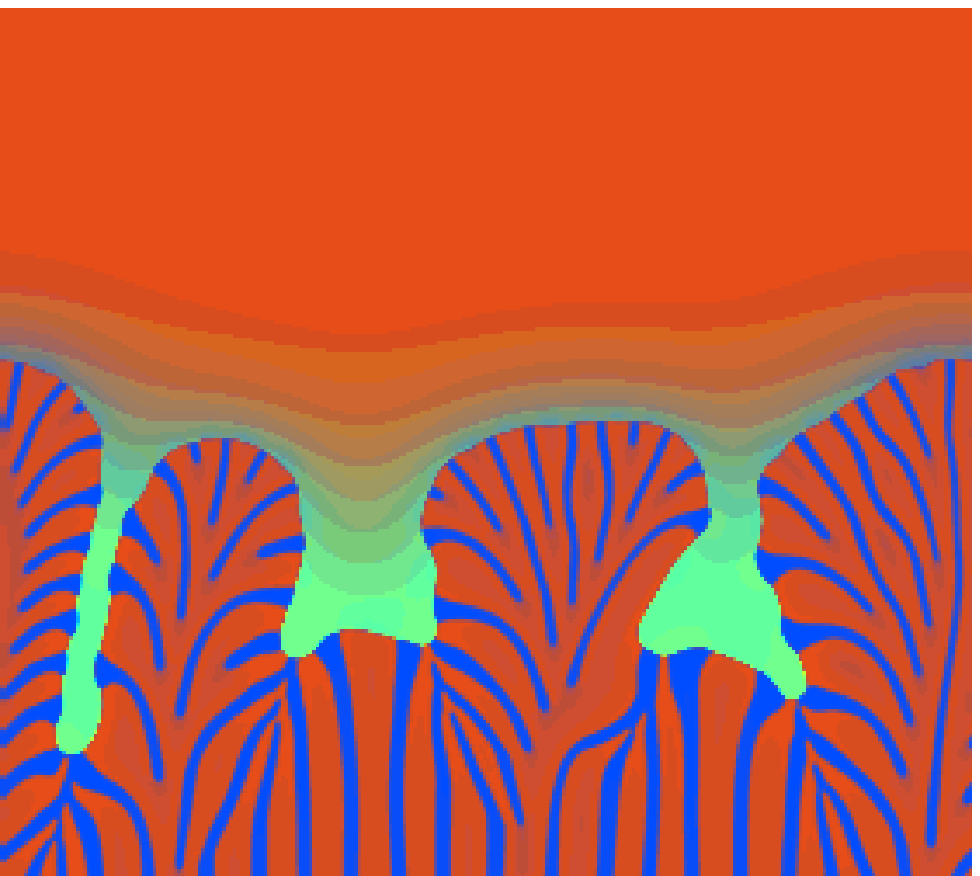,width=.32\textwidth}}
\medskip
\centerline{
  \psfig{file=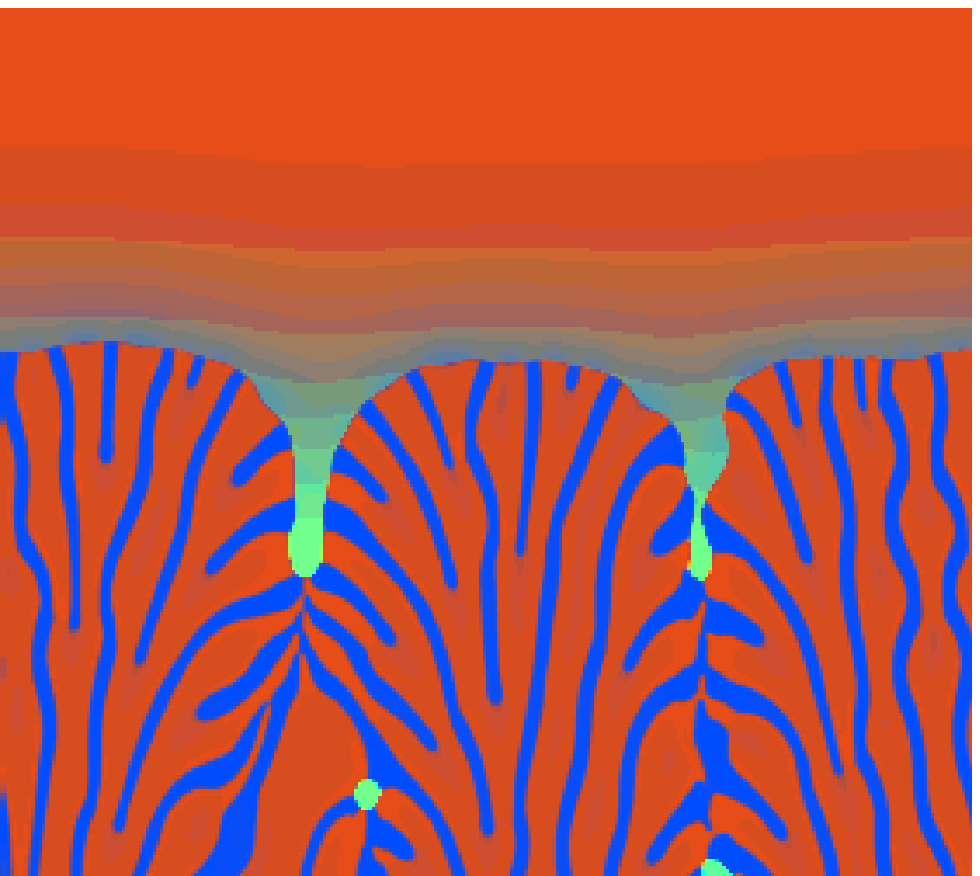,width=.32\textwidth}}
\medskip
\caption{Snapshot pictures of a run with 20 lamella pairs at 
off-eutectic composition ($u_\infty=0.3$), $g=0.1$, $w=0.2$, and
$\lambda_0/W=56$. From top to bottom: 
$t=85000, 105000, 125000, 225000$ (in units of $\tau$).}
\label{figoffeut}
\end{figure}

A quite different scenario occurs for off-eutectic
compositions. An example is shown in Fig. \ref{figoffeut}. 
The linear regime is still in good agreement with the
predictions of Ref. \cite{Plapp99}. In particular,
for sufficiently off-eutectic compositions, the 
impurity-induced long-wavelength instability competes
with the 2$\lambda$-oscillatory instability that is
already present in binary eutectics. For the
temperature gradient and impurity content chosen
in our example, the long-wavelength instability
is stationary and faster than the 2$\lambda$-O
instability. Indeed, we find that the Fourier
spectrum of the trijunction displacements is
initially dominated by the smooth long-wavelength
modes, while the 2$\lambda$-O instability
develops much more slowly. However, as soon as the 
instability becomes ``visible'', that is, the amplitude 
of the perturbation exceeds $\sim0.1\Lambda_0$,
localized finger-like structures develop
around a lamella of the minority phase and
rapidly grow ahead of the front. The fine lamellae
act almost as ``guides'' for the well-developed fingers 
during the subsequent evolution. In particular, note the long
minority lamella that is like a	``spine'' for the rightmost
finger in the third snapshot (we remind the reader that we
use periodic boundary conditions in the lateral directions;
hence, this is not a ``wall effect''). These structures,
however, are only transient. In the final stage, when the
colonies are well-developed, they have rather flat tops and
sharper ``corners'' than the fingers at eutectic composition.
In the flat parts at the center of the colonies, sometimes
a period-doubling oscillatory mode develops until it generates
some new lamellae and dies out.

Structures such as the initial localized fingers are evidently
non-linear. It thus appears that the linear regime of
the instability is much shorter for off-eutectic than
for eutectic compositions. It is presently unclear
what precisely triggers the formation of such fingers,
and under which conditions they can form.
In view of the necessary computer time, we did
not carry out a detailed study to clarify these issues.

\section{Conclusions}

We have presented a phase-field model for
eutectic solidification in the presence of ternary
impurities. This model has enabled us to carry out
large-scale simulations of colony formation starting
from arrays of up to 40 lamellae pairs. 

In the linear regime, i.e. for small perturbations of the 
unstable steady-state growth front, these simulations have
allowed us to critically test our previous linear stability
analysis \cite{Plapp99}. We find a good
overall agreement with our theoretical predictions.
Furthermore, a detailed treatment of the simulation data
has allowed us to check the assumptions made in the
linear stability analysis, and to precisely pinpoint
the reasons for the differences between the theory
and simulation results. 

The most interesting conclusion is that the growth of the lamellae is
not exactly normal to the large-scale envelope of
the composite interface, a rule originally proposed by Cahn
and used in the subsequent stability studies 
by Datye-Langer \cite{Datye81} and ourselves \cite{Plapp99}.
This effect seems to be qualitatively different
from the corrections to Cahn's rule reported previously
in other theoretical studies of binary eutectics submitted
to a strong temperature gradient \cite{Caroli90,Chen01}.
The motion of the trijunction points can be 
roughly understood as a superposition of normal
motion as stipulated by Cahn's rule and a slow
``sliding'' of the trijunctions along the front
with a sideways velocity that is proportional
to the gradient of the local lamellar spacing.
The resulting deviations of the growth angles from
$90^\circ$ are very small (below $1^\circ$);
hence, a direct measurement of this effect in
experiments is impossible, since a precise measurement of 
the growth angles is complicated by crystallographic
effects, in particular the anisotropies of the
surface tensions \cite{Akamatsu00}. However,
the growth rates of the long-wavelength modes are
very sensitive to a small change in this angle.
Studying such modes can therefore offer the
possibility to experimentally test our results.
In particular, the consequences of this
effect for the long-wavelength instability of
binary eutectics will be discussed in a
forthcoming study.

Regarding the dynamics of fully developed colonies, we find
that after the destabilization of the planar front, the
array of two-phase cells undergoes a complicated and
seemingly chaotic sequence of tip-splitting and cell
elimination events. We were unable in our simulations
to attain a steady-state configuration of the large-scale
pattern, that is, the envelope of the front. 
This result is consistent with the fact that monophase 
cellular arrays in directional solidification of dilute 
alloys are unstable
in the absence of crystalline anisotropy \cite{Kopetal,Akaetal}.
In fact, the lack of stability of the eutectic colonies
in the absence of anisotropy suggests
that the large-scale composite eutectic 
interface behaves qualitatively as a monophase front 
even beyond the linear regime. In this analogy,
the addition of solid-liquid or 
solid-solid anisotropy could potentially produce an  
effective anisotropy of the composite interface
that stabilizes 
its large-scale envelope. The quantitative exploration
of this analogy, however, is far beyond the scope of the
present work.

Regarding the comparison between our simulations
and the experimental observations of Ref.~\cite{Akamatsu00}, 
we find many similarities.
In particular, we find in the simulations the oscillatory 
unstable modes predicted by our stability analysis. Such
wavy structures are also observed in the experiments.
We also find that well-developed two-phase cells do not seem 
to reach a steady-state up to the largest times simulated.
This is in agreement with the experiments, where no steady
state has been reached even on length and time scales
far superior to the range of our simulations
(compare our Fig.~\ref{figeutspacetime} to Fig.~14 of
Ref.~\cite{Akamatsu00}).

A number of experimental observations, however, remain to be
understood. Firstly, unstable modes in the experiments
are sometimes manifested as waves that are 
emitted by localized perturbations, such as grain 
boundaries. These waves can propagate along the front, which
remains planar, rather than be a transient that precedes
colony formation. Some of these
propagating waves seem to have characteristics
of solitary waves. No such structures have been 
observed in our simulations. Furthermore, we observe
some localized two-phase fingers that play a role
during the instability of planar fronts at off-eutectic
compositions, and that are similar to structures seen in the
experiments (compare, in particular, our Fig.~\ref{figoffeut}
to Fig.~6 of Ref.~\cite{Akamatsu00}). However, other experimentally 
observed patterns, such as ``multiplet fingers'' and two-phase
dendrites are not reproduced by our simulations.
It is possible that the existence of such patterns 
depends sensitively on the structure of the eutectic phase diagram,
in particular on the asymmetry of the two solid phases and
their surface energies that have been shown to
influence the stability of binary lamellar  
eutectics \cite{Karma96}, and on crystalline anisotropy.

The present phase-field model could
easily be modified to include some degree of 
asymmetry between phases as well as both solid-liquid and
solid-solid anisotropy.
In addition, the use of more general phase-field
models with several order parameters
\cite{Wheeler96,Tiaden98,Nestler00,Lo01}, as well
as the use of more efficient phase-field
formulations \cite{Karma01} and numerical algorithms
\cite{Provatas98,Plapp00} that 
greatly enhance the accessible length and time scales,
could help to elucidate these questions in the future.
The exploration of the enormously vast parameter space 
of growth conditions and material properties that govern the
formation of complex two-phase microstructures remains, 
however, a formidable numerical task.

\acknowledgments

This research was supported by U.S. DOE Grant
No. DE-FG02-92ER45471 and benefited from
computer time at the National Energy Research
Scientific Computing Center (NERSC),
Lawrence Berkeley National Laboratory, and at
the Northeastern University Advanced Scientific
Computation Center (NU-ASCC). We thank Silv{\`e}re
Akamatsu and Gabriel Faivre for many stimulating
discussions.


\begin{thebibliography}{99}
\bibitem{rem0} Note that this definition of $u$ differs by the
  factor of $1/2$ in the denominator from the definition
  of the scaled concentration widely used in the litterature.
\bibitem{rem1} By this we understand the Helmholtz free energy,
  normalized by a typical value of the free energy density times
  the total volume of the system.
\end{thebibliography}
\end{document}